\documentclass[hidelinks,onefignum,onetabnum]{siamart251216}

\usepackage{caption}
\usepackage{subcaption}

\usepackage{amsfonts}
\usepackage{graphicx}
\usepackage{subcaption}
\usepackage{epstopdf}
\usepackage{algorithmic}
\ifpdf
  \DeclareGraphicsExtensions{.eps,.pdf,.png,.jpg}
\else
  \DeclareGraphicsExtensions{.eps}
\fi

\usepackage{amssymb}
\usepackage{physics}
\usepackage{booktabs}

\newcommand{\E}{\mathbb{E}}
\newcommand{\F}{\mathcal{F}}

\newcommand{\Q}{\mathbb{Q}}

\renewcommand{\tilde}{\widetilde}

\renewcommand{\geq}{\geqslant}
\renewcommand{\leq}{\leqslant}

\renewcommand{\bar}{\overline}
\newcommand{\ttau}{\tilde \tau}

\renewcommand{\d}{\textrm{d}}
\renewcommand{\tilde}{\widetilde}

\newsiamremark{remark}{Remark}
\newsiamremark{hypothesis}{Hypothesis}
\newsiamthm{example}{Example}
\crefname{hypothesis}{Hypothesis}{Hypotheses}
\newsiamthm{claim}{Claim}
\newsiamremark{fact}{Fact}
\crefname{fact}{Fact}{Facts}

\headers{Pricing lookback options on a quantum computer}{T. Belabbas, E. Hamel, A. MacKay, F. Paquette}

\title{Pricing lookback options on a quantum computer\thanks{Submitted to the editors DATE.
\funding{The authors acknowledge financial support from l'Autorité des Marchés Financiers, from the Pôle de recherche en finance quantique de l'Université de Sherbrooke, from the Ministère de l’Économie, de l’Innovation et de l’Énergie du Québec and from the Canada First Research Excellence Fund. Anne MacKay also acknowledges support from the Natural Science and Engineering Research Council of Canada (Grant number RGPIN-2024-05794). }}}

\author{Tania Belabbas\thanks{Quantum AlgoLab, Institut quantique, University of Sherbrooke, Sherbrooke, Quebec, Canada
  (\email{tania.belabbas@usherbrooke.ca}).}
\and Emmanuel Hamel\thanks{Autorité des marchés financiers, Quebec, Canada
  (\email{emmanuel.hamel@lautorite.qc.ca}).}
\and Anne MacKay\thanks{Department of Mathematics, University of Sherbrooke, Sherbrooke, Quebec, Canada
  (\email{anne.mackay@usherbrooke.ca}).}
\and Florence Paquette\footnotemark[4]}

\usepackage{amsopn}

\ifpdf
\hypersetup{
  pdftitle={Pricing lookback options on a quantum computer}, 
  pdfauthor={T. Belabbas, E. Hamel, A. MacKay, F. Paquette}
}
\fi

\begin{document}

\maketitle

\begin{abstract}
We develop a quantum algorithm to price discretely monitored lookback options in the Black–Scholes framework using imaginary time evolution. By rewriting the pricing PDE as a Schrödinger-type equation, the problem becomes the imaginary time evolution of a quantum state under a non-Hermitian Hamiltonian. This evolution is approximated with the Variational Quantum imaginary time evolution (VarQITE) method, which replaces the exact non-unitary dynamics with a parameterized, hardware-efficient quantum circuit. A central challenge arises from jump conditions caused by the discrete updating of the running maximum. This feature is not present in standard quantum treatments of European or Asian options. To address this, we propose two quantum-compatible formulations: (i) a sequential approach that models jumps via dedicated jump Hamiltonians applied at monitoring dates, and (ii) a simultaneous multi-function evolution that removes explicit jumps at the expense of an increased number of dimensions. We compare both approaches in terms of qubit resources, circuit complexity and numerical accuracy, and benchmark them against Monte Carlo simulations. Our results show that discretely monitored, path-dependent options with jump conditions can be handled within a variational quantum framework, paving the way toward the quantum pricing of more complex derivatives with non-smooth dynamics.

\end{abstract}



\section{Introduction}\label{seq:intro}
Quantum computing takes advantage of the quantum properties of matter to encode and process information~\cite{NielsenChuang2010}. While this idea remained theoretical for many years, recent technological advances make quantum computers a reality today. Although its capacity is still limited, quantum computing technology is improving at a tremendous rate and is expected to revolutionize a wide range of sectors, from material science to cybersecurity~\cite{Dalzell2025}. The past decade has seen remarkable progress and milestones include the demonstration of quantum utility in specific computational tasks and the continuous scaling of qubit counts and quality across various hardware modalities. The future promises fault-tolerant quantum computers, which are expected to unlock the full potential of algorithms like Shor's and Grover's. However, current research, including this work, focuses heavily on using variational quantum algorithms (VQAs) that can extract utility from today's hardware~\cite{McArdle2019,Yuan2019}, with an outlook on future possibilities. The current generation of devices, referred to as Noisy Intermediate-Scale Quantum (NISQ) machines, is characterized by a limited number of qubits and short coherence times.

Quantum computing is proposed as a potential solution to increasingly complex computations in quantitative finance~\cite{Herman2023,Orus2019}. For instance, quantum algorithms have been proposed to solve combinatorial or convex optimization problems, which have the potential to provide novel methods for efficient portfolio selection~\cite{Catalano2024,Gomez-Cadavid2024,Liu2025}. Quantum-enhanced machine learning methods have also found applications in finance~\cite{Bhasin2024,Thakkar2024,Gujju2024}.

The focus of this work is on option pricing. The pricing algorithms proposed in the quantum computing literature broadly divide into two main categories: quantum amplitude estimation (QAE) methods (sometimes called quantum Monte Carlo) and approaches based on solving pricing partial differential equations~\cite{Miyamoto2022,Fontanela2021}.

QAE-based algorithms show great potential for option pricing, but they are not adapted to the quantum machines that are currently available, as they generally call for deep circuits and a large number of qubits~\cite{Wilkens2023,Zhou2025,Fontanela2021}. An additional challenge for these methods rests in the preparation of the state, that is, the loading of the probability distribution on the quantum computer~\cite{Wilkens2023,Zhou2025}.

Although there exist different quantum methods to solve partial differential equations (PDEs), quantum time-evolution algorithms are a natural candidate for this task. The main idea is to map the pricing PDE to Schr\"{o}dinger's equation and to evolve the state of the computer's qubit register using the resulting Hamiltonian. Pricing PDEs generally yield non-Hermitian Hamiltonians, which represent an additional challenge as they prescribe non-unitary transformations. To solve this problem,~\cite{Gonzalez-Conde2023} suggest extending the Hilbert space with additionnal qubits to embed its propagator, while others suggest the use of the Schr\"{o}dingerization technique~\cite{Kumar2025}. An alternative solution, which we use in this paper, is to apply a change of variable and to consider imaginary time evolution instead, using the \texttt{VarQITE} algorithm~\cite{McArdle2019,Fontanela2021}. This algorithm prescribes the use of a parametrized circuit, whose parameters are updated following the pricing Hamiltonian.

Herein, we consider the pricing of a discretely monitored lookback option; this problem is motivated by the valuation of variable annuity contracts (see~\cite{Bacinello2011,Shevchenko2016}).
In certain cases, the financial guarantee embedded in those hybrid life insurance products can be expressed as a lookback option.
In that sense, this paper constitutes a first step towards a fully fledged quantum framework for the valuation of those products.

Our work is an extension of~\cite{Fontanela2021}, in which imaginary time evolution is used to price European and Asian options in the Black‑Scholes model. While the value functions they consider are continuous, the PDE that we solve has jump conditions stemming from the discrete monitoring of the maximum. This requires careful consideration in the design of the Hamiltonians used in the \texttt{VarQITE} algorithm.

We suggest two approaches; the first one makes use of a “jump Hamiltonian” at each jump time, while the second one relies on the simultaneous evolution of a large number of continuous functions. While the implementation of the first approach requires fewer qubits for the space discretization, the second approach yields better results, even with a limited number of qubits. We demonstrate this phenomenon in an extensive numerical illustration in Section~\ref{seq:res}.

The paper is organized as follows. In Section~\ref{seq:prob}, the financial problem and basic notions of quantum computing are introduced. Hamiltonian time evolution is described in Section~\ref{seq:ham}; this is where both methods for integrating jump conditions in the evolution are laid out. Section~\ref{seq:met} presents our methodology and Section~\ref{seq:res} contains a numerical performance analysis.

\section{Problem setup}\label{seq:prob}
In the following section, we first define the financial market dynamics and formulate the pricing problem. We also introduce essential quantum computing notions, with an emphasis on the quantum simulation of physical systems, and establish a correspondence between the option pricing PDE and the Schr\"{o}dinger equation, recasting the pricing problem into an adequate form for quantum simulation. This section sets the stage for the subsequent development of our methodology.

\subsection{Financial market}
On a probability space $(\Omega,\F,\Q)$, we define a standard Brownian motion $W = (W_\eta)_{\eta\geq 0}$ and fix a finite time horizon $T \in \mathbb R_+$. We let $\mathbb F = (\F_\eta)_{\eta\geq 0}$ denote the augmented filtration generated by $W$.  The financial market contains a risky asset, whose price process $S=(S_\eta)_{0\leq \eta \leq T}$ satisfies
\begin{align*}
    \d S_\eta = S_\eta (r \, \d \eta + \sigma \, \d W_\eta), \quad S_0 = s_0,
\end{align*}
with $r \geq 0$ and $\sigma, s_0 > 0$, as well as a risk-free bond accruing interest at the continuous rate $r$, i.e., $B_\eta = e^{r \eta}$ for $0\leq \eta \leq T$. 

Readers will recognize the Black-Scholes model, one of the simplest in the financial literature \cite{Bjork2019,Wilmott1994}, which has been generalized in various directions over the years. Our goal in this paper is to explore how quantum computing can be used to price a discretely monitored lookback option, which involves solving a PDE with jump conditions. Following \cite{Fontanela2021}, we restrict ourselves to a simple market, but we extend their work by considering a payoff that leads to a discontinuous pricing equation.

\subsection{Pricing the lookback option}  
Let $\mathcal T = \{\tilde\eta_0, \ldots, \tilde\eta_N\}$ be a uniform partition of $[0,T]$,  
and for $0 \leq \eta \leq T$, define the running maximum
\begin{align*}
    Y_\eta = \max_{s \in \mathcal T \cap [0,\eta]} S_s.
\end{align*}
The time-$T$ payoff of the lookback option is $(Y_T - S_T)^+$, where $(x)^+ = \max(0,x)$.  
The option price at time $\eta$, for an underlying price $x$ and running maximum $y$, is
\begin{align*}
    v(\eta,x,y) = \E\Big[e^{-r(T-\eta)} (Y_T - S_T)^+ \mid S_\eta = x, Y_\eta = y\Big].
\end{align*}

\begin{remark}
For notational simplicity, we include $S_\eta$ in the computation of the maximum at time $\eta$.  
This choice does not affect the payoff: including $S_T$ in $Y_T$ only impacts cases where $S_T$ exceeds the previous maximum, in which case the payoff is unchanged.  
Thus, $S_T$ may be omitted from $Y_T$ in numerical implementations.
\end{remark}

Following \cite{Wilmott1994}, we apply the transformation $v(\eta,x,y) = y \, u(\xi,z)$ with $z = x/y$ and $\xi = T-\eta$, using time to maturity $\xi$ rather than elapsed time $\eta$.  
We define $\tilde\xi_j = T - \tilde\eta_j$, so that $\mathcal T = \{\tilde\eta_0,\ldots,\tilde\eta_N\} = \{\tilde\xi_0,\ldots,\tilde\xi_N\}$.

Then, $u(\xi,z)$ solves
\begin{align}\label{eq:pricing_pde}
    -\frac{\partial u(\xi,z)}{\partial \xi} = r u(\xi,z) - r z \frac{\partial u(\xi,z)}{\partial z} - \tfrac{1}{2} \sigma^2 z^2 \frac{\partial^2 u(\xi,z)}{\partial z^2},
\end{align}
for $(\xi,z) \in ([0,T] \times (0,1)) \cup ([0,T]\setminus \mathcal T \times [1,\infty))$, with boundary conditions
\begin{align}
    &u(0,z) = (1 - z)^+,\label{eq:initial_condition}\\
    &u(\xi,0) = e^{-r\xi},\label{eq:lower_bound_condition}\\
    &\lim_{z \to \infty} \frac{z}{u(\xi,z)} \frac{\partial u(\xi,z)}{\partial z} = 1, \quad \xi \in [0,T)\setminus \mathcal T.
    \label{eq:upper_bound_condition}
\end{align}
The jump condition at monitoring dates is
\begin{equation}\label{eq:jump_condition}
    u(\xi^+,z) = z \, u(\xi^-,1), \quad (\xi,z) \in \bar{\mathcal T} \times [1,\infty),
\end{equation}
where $\bar{\mathcal T} = \mathcal T \setminus \{\tilde\xi_0, \tilde\xi_N\} = \mathcal T \setminus \{0,T\}$.  

We note that \eqref{eq:pricing_pde} holds everywhere except for $z \geq 1$ on $\mathcal T$, reflecting the fact that at a monitoring date, if $S_{\tilde\eta_i^-} > Y_{\tilde\eta_i^-}$, the running maximum resets: $Y_{\tilde\eta_i^+} = S_{\tilde\eta_i^+}$.  
This jump condition poses a challenge for quantum implementation, distinguishing this problem from those considered in \cite{Fontanela2021}.

\subsection{Review of quantum computing notions}
Our approach to solving the PDE consists in reformulating it as a Schr\"{o}dinger-type equation, where the price function is encoded as a quantum state and evolved in (imaginary) time under a suitably defined Hamiltonian using variational quantum algorithms.
For completeness, we briefly recall a minimal set of notions from quantum mechanics and quantum computing.
A more comprehensive introduction can be found in~\cite{NielsenChuang2010}.

The fundamental unit of quantum information is the qubit, the quantum analogue of the classical bit.
An $n$-qubit system is described by a state vector $\ket{\psi}$ belonging to a complex Hilbert space of dimension $2^n$.
In the computational basis, such state can be written as $
\ket{\psi} = \sum_{j=0}^{2^n-1} \alpha_j \ket{j}$,
where the complex amplitudes $\alpha_j$ satisfy the normalization condition $\sum_j |\alpha_j|^2 = 1$.
In our framework, these amplitudes are used to encode discretized values of the pricing function.

Quantum algorithms manipulate the state $\ket{\psi}$ through unitary transformations, implemented as sequences of quantum gates.
Observable quantities are extracted via measurements yielding probabilistic outcomes governed by the amplitudes of the state.
Despite this intrinsic randomness, repeated measurements (shots) allow one to estimate relevant quantities with controlled statistical error.
Additional details on quantum states, measurements, and quantum gates are provided in~\cite{NielsenChuang2010}.

\subsection{Schr\"{o}dinger equation}\label{sseq:schro}
We use the state of a quantum register to represent a vector of option prices.
The initial state of the vector corresponds to the value of the function $u$, described by \eqref{eq:pricing_pde}-\eqref{eq:jump_condition}, at $\xi = 0$ and for different values $z$.
To solve the pricing PDE, we translate it into the Schr\"{o}dinger equation, which describes the time-evolution of a quantum system, namely the qubit register in our case.
This is akin to the idea proposed by \cite{Fontanela2021}. Let $\xi = -i t$, so that \eqref{eq:pricing_pde} becomes
\begin{align}\label{eq:pricing_pde_wick_rotation}
    -i\frac{\partial u(t,z)}{\partial t} = ru(t,z) - rz\frac{\partial u(t,z)}{\partial z} - \frac 12 \sigma^2z^2 \frac{\partial^2 u(t,z)}{\partial z^2}.
\end{align}

The Schr\"{o}dinger equation describing the time-evolution of a given time-dependent quantum state $\ket{\psi(t)}$ is
\begin{align}\label{eq:schroedinger}
    i \frac{\partial\ket{\psi(t)}}{\partial t} = \hat{H}(t) \ket{\psi(t)},
\end{align}
where $\hat{H}$ is the Hamiltonian operator associated with the system. In quantum mechanics, this operator is used to describe how the system evolves in time. The PDE operator in \eqref{eq:pricing_pde_wick_rotation} can be rewritten as a Hamiltonian-like operator. Solving the PDE is then equivalent to evolving an initial state $u(0, z)$ under the dynamics generated by $\hat{H}$. 

The solution to \eqref{eq:schroedinger} is given by
    $\ket{\psi(t)} = \exp\left(-i \int_0^t \hat{H}(s) \d s\right) \ket{\psi(0)}$,
and the Hamiltonian associated with the pricing of the lookback option is thus given by the right-hand side of \eqref{eq:pricing_pde_wick_rotation}.

\section{Hamiltonian Time Evolution}\label{seq:ham}
The Hamiltonian derived in Section~\ref{sseq:schro} can be used within a time-evolution framework. 
As discussed previously, the operator appearing in the pricing PDE can be interpreted as a Hamiltonian-like operator generating the dynamics of the system.  After spatial discretization, the pricing PDE’s evolution can be written in operator form using $\hat{H}(t)$. 
Solving the discretized PDE is therefore equivalent to evolving an initial state under the action of this Hamiltonian. 
Quantum time-evolution algorithms may then be used to approximate this evolution, and suitable measurements of the resulting quantum state provide information related to the option price. 
Specifically, imaginary time evolution is well suited for this purpose because it leads to exponential propagators analogous to those appearing in the formal solution of the pricing PDE \cite{McArdle2019, Fontanela2021, Kumar2025}.

\subsection{Imaginary Time Evolution}

Consider a Wick rotation $t = -i\tau$, where $\tau \in \mathbb{R}$. 
Under this transformation, the Schr\"odinger equation becomes
\begin{align}
    \frac{\partial \ket{\psi(\tau)}}{\partial \tau}
    =
    -(\hat{H}(\tau) - E_\tau)\ket{\psi(\tau)},
    \label{eq:wick_schrod}
\end{align}
where $E_\tau = \expval{\hat{H}(\tau)}{\psi(\tau)}$ ensures normalization of the evolving state~\cite{Wick1954}. 
The corresponding formal solution can be written as
\begin{equation}\label{eq:schrod_sol}
    \ket{\psi(\tau)} = \gamma(\tau)
    \exp\left(- \int_0^\tau \hat{H}(s)\, \d s \right)
    \ket{\psi(0)},
\end{equation}
where $\gamma(\tau)$ is a normalization factor for the state $\ket{\psi(\tau)}$~\cite{McArdle2019}. 
The exponential operator is non-unitary and generates imaginary time propagation.

Because this exponential operator has the same structure as the propagator appearing in the pricing PDE, imaginary time evolution provides a natural framework for implementing the propagation of the option value on quantum hardware. 
Moreover, imaginary time evolution can be implemented variationally, allowing the dynamics of $\ket{\psi(\tau)}$ to be approximated through differential equations governing the parameters of a parameterized quantum circuit~\cite{Yuan2019, McArdle2019}.

\subsection{Application to the lookback option}

The pricing PDE of the lookback option given by \eqref{eq:pricing_pde} - \eqref{eq:jump_condition} describes a piecewise continuous function with jumps occurring at times $\tilde \tau_j$, $j=1,\ldots,N-1$.

When solving the equation classically using finite difference methods, the jump condition can be treated separately by \textit{pausing} the continuous evolution and adjusting the function value at each jump time. With imaginary time evolution, the Hamiltonian describing the dynamics of the price must be carefully designed. We have identified two ways to do so. Our first method consists in alternating between two operators: a Hamiltonian describing the continuous evolution, used between jump times, and another one simulating the jump dynamics, applied at jump times, see Section \ref{sec:m1}. Second, we consider a system of $N$ PDEs in which every equation is continuous. $N-1$ PDEs are linked to the value of the function $u(\tau, z)$ for $z \geq 1$ after each of the jumps, at times $\tilde \tau_{N-1},\ldots,\tilde \tau_{1}$. Details of this second method can be found in section \ref{sec:m2}.

\subsubsection{Method 1: Jump Hamiltonian}\label{sec:m1}

To implement the time-evolution on a quantum computer, we choose a uniform discretization of the half-line $[0,\infty)$ on $n$ points, $\vec z= \{z_0,\ldots,z_{n-1}\}$, with
$z_0 = 0$, $z_{n-1} = z_{max}$ and $\Delta_z = z_i - z_{i-1},$ $i=1,\ldots,n-1$.
The state of the qubit register encodes the solution of the pricing PDE (up to the normalization term $\gamma(\tau)$) at each discretization point $z_i$, $i=0,\ldots,n-1$; the amplitude of each basis state gives the solution for that specific $z_i$, so that
\begin{align*}
    \ket{\psi(\tau)} = \gamma(\tau) u(\tau,\vec z),
\end{align*}
for $\tau \in [0,T]$, where $\gamma(\tau)$ represents the normalization term induced by imaginary time evolution (see the algorithms in Section \ref{seq:met} for more details on its computation). 

To evolve the initial state, we define two different time-evolution matrix operators, $\hat H_C$ and $\hat H_J$ using the right-hand side of \eqref{eq:pricing_pde_wick_rotation} and the jump condition \eqref{eq:jump_condition}.
$\hat H_C$ reflects the continuous evolutions of $\ket{\psi(\tau)}$ for $\tau \in [0,\tilde \tau_1 - h] \cup (\tilde \tau_1, \tilde \tau_2 -h] \cup \ldots \cup (\tilde \tau_{N-1}, T]$, and $\hat H_J$ includes the jump condition and describes the evolution of $\ket{\psi(\tau)}$ when $\tau \in (\tilde\tau_1-h, \tilde\tau_1]\cup\ldots\cup(\tilde\tau_{N-1}-h, \tilde\tau_{N-1}]$, with $h > 0$ small.
Thus, the general Hamiltonian $\hat H_{M1}: [0,T] \mapsto \mathbb R^{n\times n}$ is time-dependent and given by
\begin{align*}
    \hat H_{M1}(\tau) =
    \begin{cases}
        \hat H_J, & \tau \in (\tilde \tau_1-h, \tilde \tau_1]\cup\ldots\cup(\tilde \tau_{N-1}-h, \tilde \tau_{N-1}]\\
        \hat H_C, & \text{otherwise.}
    \end{cases}
\end{align*}

$\hat H_C$ is obtained directly by discretizing the right-hand side of \eqref{eq:pricing_pde_wick_rotation} via second-order finite differences, yielding
\begin{align}
	\frac{\partial u(\tau,z_i)}{\partial \tau}
    &= \alpha_i u(\tau,z_{i-1}) + \beta_i u(\tau,z_i) + \gamma_i u(\tau,z_{i+1}),
    \label{eq:H1_all}
\end{align}
with

\begin{align}
    \alpha_i = -\frac{rz_i}{2\Delta_z} + \frac{\sigma^2 z_i^2}{2{\Delta_z}^2},
    \qquad
    \beta_i = -r-\frac{\sigma^2 z_i^2}{\Delta_z^2},
    \qquad
    \gamma_i = \frac{rz_i}{2\Delta_z} +\frac{\sigma^2 z_i^2}{2 {\Delta_z}^2}.
    \label{eq:alpha_beta_gamma}
\end{align}

The upper and lower boundary conditions are also discretized and expressed in Hamiltonian form, so that from \eqref{eq:lower_bound_condition}, we have
\begin{align}
    \frac{\partial u(\tau,z_0)}{\partial \tau} = - r u(\tau,z_0).
    \label{eq:H1_lower}
\end{align}
Discretizing the upper boundary condition \eqref{eq:upper_bound_condition} using a backward finite difference in $z$ gives $u(\tau, z_{n-1}) = \frac{z_{n-1}}{z_{n-2}} u(\tau,z_{n-2})$, from which we get
\begin{align}
    \frac{\partial u(\tau,z_{n-1})}{\partial \tau} = \frac{z_{n-1}}{z_{n-2}} \left(\alpha_{n-2} u(\tau,z_{n-3}) + \beta_{n-2}u(\tau,z_{n-2}) + \gamma_{n-2} u(\tau,z_{n-1})\right).
    \label{eq:H1_upper}
\end{align}

From \eqref{eq:H1_all}, \eqref{eq:H1_lower} and \eqref{eq:H1_upper}, we write $\hat H_C$ as
\begin{equation}
\label{eq:matrix_HC}
	\hat H_C =
	\begin{bmatrix}
		r & 0 & 0 & 0 & \ldots & 0 & 0 & 0 \\
		\alpha_1 & \beta_1 & \gamma_1 & 0 & \ldots & 0 & 0 & 0 \\
		0 & \alpha_2 & \beta_2 & \gamma_2  & \ldots & 0 & 0 & 0 \\
		\vdots & \vdots & \vdots & \vdots &\ddots & \vdots & \vdots & \vdots \\
		0 & 0 & 0 & 0 & \ldots & \alpha_{n-2} & \beta_{n-2} & \gamma_{n-2} \\
		0 & 0 & 0 & 0 & \ldots & \frac{z_{n-1}}{z_{n-2}}\alpha_{n-2} & \frac{z_{n-1}}{z_{n-2}}\beta_{n-2} & \frac{z_{n-1}}{z_{n-2}}\gamma_{n-2} \\
	\end{bmatrix}.
\end{equation}

The matrix operator $\hat H_J$ describes the evolution of the function $u$ from $\tilde \tau_j - h$ to $\tilde \tau_j^+$, $j = 1, \ldots, N-1$.
For the discussion below, we fix $j \in \{1,\ldots,N-1\}$.
Here, $u(\tilde \tau_j^+, z)$ denotes the value of the function after the jump.
Because of the jump condition \eqref{eq:jump_condition}, the time-derivative of $u$ does not exist at $\tilde\tau_j$ for $z_i > 1$, so instead we use the finite difference $\frac{1}{h}(u(\tilde\tau_j^+,z_i)-u(\tilde\tau_j-h,z_i))$ to define $\hat H_J$.
Going forward, to simplify the notation, we implicitly define $\lambda$ by $z_\lambda = 1$.

To derive the correct form for $\hat H_J$, we express the finite difference $\frac{1}{h}(u(\tilde\tau_j^+,z_i)-u(\tilde\tau_j-h,z_i))$ in terms of the vector $(u(\ttau_j - h, z_0), \ldots, u(\ttau_j - h, z_{n-1}))^\top$.
For $j=1,\ldots,\lambda$, the function $u$ is continuous at $\ttau_j$, and thus \eqref{eq:H1_all} holds.
For $i=\lambda+1,\ldots,n+1$, using the jump condition \eqref{eq:jump_condition}, we observe that
\begin{align}
    \frac{u(\ttau_j^+,z_i) - u(\ttau_j-h,z_i)}{h} =
    \frac{z_i u(\ttau_j^-,z_\lambda) - u(\ttau_j-h,z_i)}{h}.
    \label{eq:findiff_method1}
\end{align}
Since $u$ is continuous on $[\ttau_j-h,\ttau_j)$, we use a first-order approximation of the time derivative in $\eqref{eq:pricing_pde}$ to write
\begin{equation}
    \begin{split}
        u(\ttau_j^-,z_\lambda) &\approx u(\ttau_j - h, z_\lambda)\\
    &\quad+h \left(ru(\ttau_j-h,z_\lambda)-r \frac{\partial u(\ttau_j-h,z_\lambda)}{\partial z} - \frac{\sigma^2}{2}\frac{\partial^2 u(\ttau_j-h,z_\lambda)}{\partial z^2}\right).
    \end{split}
    \label{eq:utau_method1}
\end{equation}
Approximating the derivatives in \eqref{eq:utau_method1} with central finite differences and using the resulting expression in \eqref{eq:findiff_method1} yields
\begin{align*}
    \frac{u(\ttau_j^+,z_i) - u(\ttau_j-h,z_i)}{h} &=
    a_i u(\ttau_j-h,z_{\lambda-1}) + b_i u(\ttau_j-h,z_{\lambda-1})\\
    & \quad + c_i u(\ttau_j-h,z_{\lambda+1}) - \frac 1h u(\ttau_j-h,z_i),
\end{align*}
where
\begin{equation*}
    a_i = \frac{z_i}{2\Delta_z} \left(r - \frac{\sigma^2}{\Delta_z}\right),
    \qquad
    b_i = z_i \left(\frac 1h+r+\frac{\sigma^2}{\Delta_z^2}\right)
    \qquad
    c_i = -\frac{z_i}{2\Delta_z} \left(r + \frac{\sigma^2}{\Delta_z}\right).
\end{equation*}

The resulting time-evolution operator $\hat H_J$ is given by
\begin{equation*}
    \hat H_J =
    \begin{bmatrix}
    r & 0 & 0 & 0 & \ldots & 0 & 0 & 0 & 0 & \ldots & 0\\
    \alpha_1 & \beta_1 & \gamma_1 & 0 & \ldots & 0 & 0 & 0 & 0 & \ldots & 0\\
    0 & \alpha_2 & \beta_2 & \gamma_2 & \ldots & 0 & 0 & 0 & 0 & \ldots & 0\\
    \vdots & \vdots & \vdots & \vdots & \ddots & \vdots & \vdots & \vdots & \vdots &  & \vdots\\
    0 & 0 & 0 & 0 & \ldots & \alpha_\lambda & \beta_\lambda & \gamma_\lambda & 0 & \ldots & 0\\
    0 & 0 & 0 & 0 & \ldots & a_{\lambda+1} & b_{\lambda+1} & c_{\lambda+1} + \frac 1h & 0 & \ldots & 0\\
    0 & 0 & 0 & 0 & \ldots & a_{\lambda+2} & b_{\lambda+2} & c_{\lambda+2} &  \frac 1h & \ldots & 0\\
    \vdots & \vdots & \vdots & \vdots &  & \vdots & \vdots & \vdots & \vdots & \ddots & \vdots \\
    0 & 0 & 0 & 0 & \ldots & a_{n-1} & b_{n-1} & c_{n-1} & 0 & \ldots & \frac 1h\\
    \end{bmatrix}.
\end{equation*}

We remark that $\hat H_J$ depends on $h$, the size of the time-step on which we apply the operator.
This will affect the quantum algorithm we develop for its implementation.

\subsubsection{Method 2: Continuous functions}\label{sec:m2}

In this second method, we consider a system of $T$ continuous PDEs.

\paragraph{Case $\mathbf{N=2}$}

To facilitate the presentation of this method, we first give an example when $N=2$, that is, when the maximum is computed over 2 points only. To simplify the presentation, we assume w.l.o.g. that $T=2$ and $\bar{\mathcal T} = \{1\}$; the jump condition only applies at one time point.
The jump condition, defined in \eqref{eq:jump_condition}, can be written in this case as
\begin{align*}
    u(1^+,z) =
    \begin{cases}
    u(1^-,z), &z < 1\\
    zu(1^-,1), &z \geq 1.
    \end{cases}
\end{align*}
Heuristically, the idea of this method is to ``evolve the jump value'' $zu(1,1)$ from $\tau = 0$ to $\tau = 1$ and to incorporate this value in the function $u$ at $\tau = 1$ for $z \geq 1$.
To do so, we consider two functions, $u_i: [0,1]\times (0,\infty) \mapsto \mathbb R_+$, $i=1,2$, with $u_1$ defined by \eqref{eq:pricing_pde} to \eqref{eq:upper_bound_condition}, and
\begin{align*}
    u_2(\tau,z) = zu_1(\tau,1),
\end{align*}
That is, for $\tau \in [0,1]$, $u_2$ satisfies
\begin{align}
    -\frac{\partial u_2(\tau,z)}{\partial \tau} = rzu(\tau,1) - rz\frac{\partial u(\tau,z)}{\partial z}\bigg\rvert_{z=1} - \tfrac{1}{2} \sigma^2 z \frac{\partial^2 u(\tau,z)}{\partial z^2}\bigg\rvert_{z=1},
    \label{eq:u2_pde}
\end{align}
with boundary conditions $u_2(0,z) = 0$ and $u_2(\tau,0) = 0$.

Then, on $[0,1)$, $u(\tau,z) = u_1(\tau,z)$ for $z > 0$.
At $\tau = 1^+$ (after the jump), following \eqref{eq:jump_condition}, we have $u(1^+,z) = u_1(1,z)$ for $z \in (0,1)$ and $u(1^+,z) = u_2(1,z)$ for $z \in [1,\infty)$, and for $\tau \in (1,2]$, $u$ solves \eqref{eq:pricing_pde}, \eqref{eq:lower_bound_condition} and \eqref{eq:upper_bound_condition}.
Using functions $u_1$ and $u_2$ for $\tau \in [0,1]$ allows us to express $u$ without explicitly using the jump condition.

With this representation, it is possible to implement the time-evolution of $u$ on a quantum computer without the need for jump operators.
However, encoding both functions $u_1$ and $u_2$ requires an additional qubit to double the size of the state vector.

We use the same discretization $(z_i)_{i=0}^{n-1}$ as in Method 1, which means that $(m+1)$ qubits are needed to obtain a system with $2^{m+1}=2n$ basis states.
Initially, the amplitudes of the first $n$ states encode the values of $u_1$ and those of the last $n$ states, $u_2$. In other words, the state $\ket{\varphi(\tau)}$ represents
\begin{align*}
    \ket{\varphi(\tau)} =
    \gamma(\tau)
    (u_1(\tau,z_0),\ldots,u_1(\tau,z_{n-1}),u_2(\tau,z_0),\ldots,u_2(\tau,z_{n-1}))^\top,
\end{align*}
for $\tau \in (0,1)$.
We also recall that $\lambda$ is implicitly defined by $z_\lambda = 1$.

The resulting Hamiltonian $\hat H_{M2}$ is time-dependent.
For $\tau \in [0,1)$, $\hat H_{M2}$ represents the continuous evolution of $u_1$ and $u_2$ separately and is given by
\begin{equation*}
\hat H_{M2} (\tau) =
\begin{bmatrix}
    \hat H_C & 0_{n\times n} \\
    \hat H_U & 0_{n \times n} \\
\end{bmatrix},
\end{equation*}
where $\hat H_C$, defined in \eqref{eq:matrix_HC}, describes the continuous evolution of $u_1$, $0_{n\times n}$ is a $n \times n$ null matrix.
$\hat H_U$ comes from a first-order discretization of the $z$-derivatives in \eqref{eq:u2_pde}.
It is a $n \times n$ matrix in which columns $\lambda-1$, $\lambda$ and $\lambda+1$ are the only non-zero columns.
We write $\hat H_U = \hat H_{UW} + \hat H_{UE}$, with
\begin{equation*}
    \hat H_{UW} =
    \begin{bmatrix}
        0 & \ldots & 0& z_0 \alpha_\lambda & z_0 \beta_\lambda &  0 & \ldots & 0\\
        0 & \ldots & 0 & z_1 \alpha_\lambda & z_1 \beta_\lambda &  0 & \ldots & 0\\
        \vdots & & \vdots & \vdots & \vdots & \vdots & & \vdots \\
        0 & \ldots & 0 & z_{n-1} \alpha_\lambda & z_{n-1} \beta_\lambda & 0 & \ldots & 0\\
    \end{bmatrix},
\end{equation*}
where the non-zero columns are columns $\lambda - 1$ and $\lambda$, and
\begin{equation}
    \hat H_{UE} =
    \begin{bmatrix}
        0 & \ldots & 0 & z_0 \gamma_\lambda & 0 & \ldots & 0\\
        0 & \ldots & 0 & z_1 \gamma_\lambda & 0 & \ldots & 0\\
        \vdots & & \vdots & \vdots & \vdots & \vdots & \vdots\\
        0 & \ldots & 0 & z_{n-1} \gamma_\lambda & 0 & \ldots & 0\\
    \end{bmatrix}
\end{equation}
where the only non-zero column is column $\lambda + 1$. In the above, $\alpha_\lambda$, $\beta_\lambda$ and $\gamma_\lambda$ are defined by \eqref{eq:alpha_beta_gamma}.
The reason for the decomposition of $\hat H_U$ into $\hat H_{UW}$ and $\hat H_{UE}$ will be clear when we consider a general maturity $T$.

For $\tau \in [1,2]$, $\hat H_{M2}$ represents the evolution of $u$, whose values are encoded in the first $\lambda$ and the last $n-\lambda$ elements of $\ket{\varphi(\tau)}$.
Thus, the Hamiltonian needs to act on these specific states only and represents the same evolution as $\hat H_C$ does.
The resulting operator for $\tau \in [1,2]$ is given by
\begin{equation}
    \hat H_{M2}(\tau) =
    \begin{bmatrix}
        \hat H_{CNW} & \hat H_{CNE}\\
        \hat H_{CSW} & \hat H_{CSE}
    \end{bmatrix},
\end{equation}
where $\hat H_{CNW}$, $\hat H_{CSE}$, $\hat H_{CNE}$, and $\hat H_{CSW}$ are the block matrices of size $n \times n$ described below.
We remark that the index notation ``NW'', ``SE'', ``NE'' and ``SW'' describes their position in the matrix $\hat H_{M2}$ using cardinal directions.
$\hat H_{CNW}$ governs the continuous evolution of the first $\lambda + 1$ entries of $\ket{\varphi(\tau)}$, representing the (normalized) value of $u$ for $z_i \leq 1$.
It is given by the first $\lambda + 1$ rows of the matrix $\hat H_C$, with the element in position $(\lambda+1, \lambda + 2)$ replaced by 0, that is
\begin{equation*}
\hat H_{CNW} = 
    \begin{bmatrix}
        r & 0 & 0 & 0 & \ldots & 0 & 0 & 0 & \ldots & 0 \\
        \alpha_1 & \beta_1 & \gamma_1 & 0 & \ldots & 0 & 0 & 0 & \ldots & 0 \\
        0 & \alpha_2 & \beta_2 & \gamma_2 & \ldots & 0 & 0 & 0 & \ldots & 0 \\
        \vdots & & \ddots & \ddots & \ddots & \vdots & \vdots & \vdots & \ddots & \vdots \\
        0 & 0 & 0 & 0 & \ldots &\alpha_\lambda & \beta_\lambda & 0 & \ldots & 0\\
        0 & 0 & 0 & 0 & \ldots & 0 & 0 & 0 & \ldots & 0\\
        \vdots & \vdots & \vdots & \vdots & \ddots & \vdots & \vdots & \vdots & \ddots & \vdots\\
        0 & 0 & 0 & 0 & \ldots & 0 & 0 & 0 & \ldots & 0\\
    \end{bmatrix}.
\end{equation*}
$\hat H_{CSE}$ represents the evolution of the last $n-\lambda-1$ elements of $\ket{\varphi(\tau)}$, reflecting the values of $u$ for $z > 1$.
It is given by the last $n - \lambda - 1$ rows of matrix $\hat H_C$, with the element in position $(\lambda+2, \lambda)$ replaced by 0, so that
\begin{equation*}
\hat H_{CSE} =
    \begin{bmatrix}
        0 & \ldots & 0 & 0 & 0 & 0 &\ldots & 0 & 0 & 0\\
        \vdots & \ddots & \vdots & \vdots & \vdots & \vdots & & \vdots & \vdots & \vdots \\
        0 & \ldots & 0 & 0 & 0 & 0 &\ldots & 0 & 0 & 0\\
        0 & \ldots & 0 & \beta_{\lambda+ 1} & \gamma_{\lambda+1} & 0 & \ldots & 0 & 0 & 0\\
        0 & \ldots & 0 & \alpha_{\lambda+2} & \beta_{\lambda+2} & \gamma_{\lambda+2} & \ldots & 0 & 0 & 0\\
        \vdots & \ddots & \vdots & &\ddots & \ddots & \ddots & \vdots & \vdots & \vdots\\
        0 & \ldots & 0 & 0 & 0 & 0 & \ldots & \alpha_{n-2} & \beta_{n-2} & \gamma_{n-2}\\
        0 & \ldots & 0 & 0 & 0 & 0 & \ldots & \frac{z_{n-1}}{z_{n-2}} \alpha_{n-2} & \frac{z_{n-1}}{z_{n-2}} \beta_{n-2} & \frac{z_{n-1}}{z_{n-2}} \gamma_{n-2}
    \end{bmatrix}
\end{equation*}
Matrices $\hat H_{CSW}$ and $\hat H_{CNE}$ represent the interactions between the first $n$ and the last $n$ elements of the state vector.
They are filled with zero, except for one element each; the element in position $(\lambda+2,\lambda)$ of matrix $\hat H_{CSW}$ is equal to $\alpha_{\lambda + 1}$, and the element in position $(\lambda+1,\lambda+2)$ of matrix $\hat H_{CNE}$ is $\gamma_\lambda$. Note that
\begin{equation*}
    \hat H_C = \hat H_{CNW} + \hat H_{CSE} + \hat H_{CNE} + \hat H_{CSW},
\end{equation*}
which reflects the fact that together, matrices $\hat H_{CNW}$, $ \hat H_{CSE}$, $\hat H_{CNE}$ and $\hat H_{CSW}$ evolve specific elements of $\ket{\varphi(\tau)}$ according to $\hat H_C$.

\paragraph{General case}

We generalize this idea to the case $N \in \mathbb N$, $N \geq 2$, that is, $\bar {\mathcal T} = \{\ttau_1,\ldots,\ttau_{N-1}\}$; the maximum is computed over more than two points.
To simplify the presentation, we assume that $T \in \mathbb N$.  
We still consider functions $u_1$ and $u_2$, which we defined previously, and define functions $u_j: [0,\ttau_{j-1}] \times (0,\infty) \mapsto \mathbb R_+$ for $j \in \{3,\ldots,N\}$.
As before, for $\tau \in [0,\ttau_1)$, $u(\tau,z) = u_1(\tau,z)$, and
\begin{equation*}
    u(\ttau_1^+,z) =
    \begin{cases}
        u_1(\ttau_1,z), & z \in (0,1)\\
        u_2(\ttau_1,z), & z \in [1,\infty).
    \end{cases}
\end{equation*}
In general, for $\tau \in (1,T]\setminus \bar{\mathcal{T}}$, $u$ satisfies \eqref{eq:pricing_pde}, \eqref{eq:lower_bound_condition} and \eqref{eq:upper_bound_condition}, and
\begin{align*}
    u_j(\tau,z) =
    \begin{cases}
        zu_1(\tau,1), & \tau \in [0,\tilde \tau_1]\\
        zu(\tau,1), & \tau \in (\tilde \tau_1,\tilde \tau_{j-1}]
    \end{cases}
\end{align*}
The jump condition is reflected by
\begin{equation*}
    u(\ttau_j^+,z) =
    \begin{cases}
        u(\ttau_j,z), & z \in (0,\tilde 1)\\
        u_{j+1}(\ttau_j,z), & z \in [1,\infty).
    \end{cases}
\end{equation*}
The definition is recursive: $u$ is defined on $[0,2]$ using $u_1$ and $u_2$, which allows the definition of $u_3$ on $[0,2]$.
Then, $u_3$ is used to specify $u(\tau_2^+,z)$, and \eqref{eq:pricing_pde}, \eqref{eq:lower_bound_condition} and \eqref{eq:upper_bound_condition} specify $u$ up to $\ttau_3$, and so on.

To implement this method on a quantum computer, we need a vector $\ket{\varphi(\tau)}$ of size $nN$.
For $\tau \in [0,1)$, the first $n$ entries represent $u_1(\tau,z_0), \ldots, u_1(\tau,z_{n-1})$, and entries $(k-1)n+1$ to $kn$ represent the values $u_k(\tau,z_0), \ldots, u_k(\tau,z_{n-1})$, $k \in \{2,\ldots,N\}$.
Thereafter, for $\tau \in [\ttau_j,\ttau_{j+1})$, the first $\lambda$ elements of $\ket{\varphi(\tau)}$ are associated with $u(\tau,z_0), \ldots, u(\tau,z_\lambda)$, and the elements $(j-1)n+\lambda+2$ to $jn$ represent values $u(\tau,z_{\lambda+1}),\ldots,u(\tau,z_{n-1})$.

The resulting Hamiltonian is a matrix of size $nN \times nN$ and is time-dependent.
For $\tau \in [0,\ttau_1)$, $\hat H_{M2}$ has its first $n$ columns filled with matrices $\hat H_C$ and $\hat H_U$. That is,
\begin{equation*}
    \hat H_{M2}(\tau) =
    \begin{bmatrix}
        \hat H_C & 0_{n \times n} & \ldots & 0_{n \times n} \\
        \hat H_U & 0_{n \times n} & \ldots & 0_{n \times n} \\
        \vdots & \vdots & & \vdots \\
        \hat H_U & 0_{n \times n} & \ldots & 0_{n \times n} \\
    \end{bmatrix}.
\end{equation*}

For $\tau \in [\ttau_j,\ttau_{j+1})$, $j \in \{1,\ldots,N-1\}$, the first $(j+1)n$ rows represent the evolution of the function $u$ via the decomposition of $\hat H_C$ into matrices $\hat H_{CSE}$, $\hat H_{CNW}$, $\hat H_{CSW}$ and $\hat H_{CNE}$.
The last $(N-j-1)n$ rows govern the evolution of functions $u_{j+2}, \ldots, u_{N}$ through matrices $\hat H_{UW}$ and $\hat H_{UE}$.
Then, for $\tau \in [\ttau_j, \ttau_{j+1})$, the Hamiltonian is given by
\begin{equation}
\label{eq:matrix_HM2}
    \hat H_{M2}(\tau) =
    \begin{bmatrix}
        \hat H_{CNW} & 0_{n \times n} & \ldots & 0_{n \times n} & \hat H_{CNE} & 0_{n \times n} & \ldots & 0_{n \times n} \\
        0_{n \times n} & 0_{n \times n} & \ldots & 0_{n \times n} & 0_{n \times n} & 0_{n \times n} & \ldots & 0_{n \times n} \\
        \vdots & \vdots & & \vdots & \vdots & \vdots & & \vdots \\
        0_{n \times n} & 0_{n \times n} & \ldots & 0_{n \times n} & 0_{n \times n} & 0_{n \times n} & \ldots & 0_{n \times n} \\
        \hat H_{CSW} & 0_{n \times n} & \ldots & 0_{n \times n} & \hat H_{CSE} & 0_{n \times n} & \ldots & 0_{n \times n} \\
        \hat H_{UW} & 0_{n \times n} & \ldots & 0_{n \times n} & \hat H_{UE} & 0_{n \times n} & \ldots & 0_{n \times n} \\
        \vdots & \vdots & & \vdots & \vdots & \vdots & & \vdots \\
        \hat H_{UW} & 0_{n \times n} & \ldots & 0_{n \times n} & \hat H_{UE} & 0_{n \times n} & \ldots & 0_{n \times n} \\
    \end{bmatrix},
\end{equation}
where $\hat C_{H}$ starts at column $j\times n + 1$, and at row $j\times n + 1$.
For $\tau \in [\ttau_{N-1},\ttau_N]$, the Hamiltonian is given by
\begin{equation*}
    \hat H_{M2}(\tau) =
    \begin{bmatrix}
        \hat H_{CNW} & 0_{n\times n} & \ldots & 0_{n\times n} & \hat H_{CNE} \\
        0_{n\times n} & 0_{n\times n} & \ldots & 0_{n\times n} & 0_{n\times n} \\
        \vdots & \vdots & & \vdots & \vdots \\
        0_{n\times n} & 0_{n\times n} & \ldots & 0_{n\times n} & 0_{n\times n} \\
        \hat H_{CSW} & 0_{n \times n} & \ldots & 0_{n \times n} & \hat C_{H} \\
    \end{bmatrix}.
\end{equation*}

This second method requires more qubits than method 1, and computational resources increase with the number of monitoring dates $N$.
For $N$ monitoring dates and $n$ discretization points in space, a state vector of size $nN$ is needed.
If $n=2^m$, then $\lceil\log_2 N \rceil + m$ qubits are needed to implement this method.
Then, as $N$ grows, the number of qubits increases at the rate $\log_2 N$.
However, we note that for $N \rightarrow \infty$, the maximum is continuously monitored and the function $u$ becomes continuous, and the implementation of the algorithm is simplified (see \cite{Fontanela2021}).

We also remark that when $\log_2 N \notin \mathbb N$, only the first $nN$ elements of the state vector will be used for method 2.
The remaining elements should simply be set to $0$.
The Hamiltonian that governs the entire state vector has size $n\times 2^{\lceil \log_2 N \rceil}$.
In each dimension, the first $nN$ elements are represented by matrix $\hat H_{M2}$; the rest of the matrix is filled with zeros.

\section{Methodology}\label{seq:met}
As mentioned in Section \ref{seq:ham}, pricing a financial derivative requires calculating the evolution of a quantum state $\ket{\psi}$ under a specific, often non-unitary propagation. The problem of option pricing can be mapped to finding the state after imaginary time evolution and the formal evolution from an initial state $\ket{\psi(0)}$ to a state at imaginary time $\tau$, $\ket{\psi(\tau)}$, is given by:
\begin{align}
    \ket{\psi(\tau)} = \gamma(\tau) \exp\left(-\int_0^\tau \hat{H}(s) \, \d s\right)\ket{\psi(0)}.
\end{align}
We have developed two algorithms to solve this problem. The non-unitary nature of the exponential operator implies that its direct, exact simulation is incompatible with the unitary operations native to quantum hardware. To address this, we employ the Variational Quantum imaginary time evolution (\texttt{VarQITE}) method. In both algorithms, the Hamiltonians $\hat H_C$, $\hat H_J$ (in Algorithm \ref{alg:sequential_evolution}), and $(\hat H_j)_{j=1}^N$ (in Algorithm \ref{alg:effective_hamiltonians}) are treated as time-independent within each evolution step.

\subsection{Algorithm \ref{alg:sequential_evolution}: Jump Hamiltonian} 

The first algorithm proposed, Algorithm~\ref{alg:sequential_evolution}, incorporates discrete revaluation jumps as fixed-interval perturbations within the imaginary time evolution. This approach naturally reflects the sequential, period-by-period structure of the segregated fund with annual revaluation. Algorithm~\ref{alg:sequential_evolution} evolves under $\hat H_C$ for continuous dynamics and $\hat H_J$ for jumps.

\begin{algorithm}
    \caption{Sequential Imaginary Time Evolution with Periodic Jumps}
    \label{alg:sequential_evolution}
    \begin{algorithmic}[H]
        \REQUIRE Rate $r$, Volatility $\sigma$, Maturity $T$, Discretization $\vec{z}$, Revaluation epochs $\ttau_j$, Jump period size $h$, $\bar{\mathcal{T}}=\{1,\ldots,N-1\}$
        \ENSURE Option price $u(0, \vec{z})$
        \STATE Construct Evolution Hamiltonian $\hat{H}_{M1}(\tau)$ with $\hat{H}_{C}$ and jump Hamiltonian $\hat{H}_{J}$
        \STATE Prepare initial state $|\psi(\tau)\rangle \leftarrow u(0, \vec{z})$
        \STATE Set current time $\tau \leftarrow 0$ (Imaginary time)
        \FOR{$j \in \bar{\mathcal{T}}$}
            \STATE Evolve $|\psi(\tau)\rangle$ to $|\psi(\ttau_j - h)\rangle$ using $\hat{H}_{C}$ via chosen method (here: \texttt{VarQITE})
            \STATE Evolve $|\psi(\ttau_j - h)\rangle$ to $|\psi(\ttau_j)\rangle$ using $\hat{H}_{J}$ via chosen method (here: \texttt{VarQITE}).
            \STATE $|\psi(\tau)\rangle \gets |\psi(\ttau_j)\rangle$
            \STATE $\tau \gets \ttau_i$
        \ENDFOR
        \STATE Evolve $|\psi(\tau)\rangle$ to $|\psi(\tau_{N})\rangle$ using $\hat{H}_{C}$ via chosen method (here: \texttt{VarQITE})
        \STATE Extract and adjust price from final state: 
        $u(\tau_N, \vec{z}) \gets \frac{e^{-rT}}{\langle 0| \psi(\tau_N) \rangle} \cdot |\psi(\tau_{N})\rangle$
        \RETURN $u(\tau_N, \vec{z})$
    \end{algorithmic}
\end{algorithm}

\subsection{Algorithm \ref{alg:effective_hamiltonians}: Effective Time-Period Hamiltonians}

Algorithm~\ref{alg:effective_hamiltonians} implies pre-calculating a distinct Hamiltonian $\hat{H}_j = \hat H_{M2}(\tau), \, \tau \in [\tilde \tau_{j-1}, \tilde \tau_j)$ for each period $j\in \{1,\ldots,N-1\}$ following \eqref{eq:matrix_HM2}. This approach consolidates the effects of both the continuous dynamics and the boundary conditions into a single imaginary time evolution process for each interval. Algorithm~\ref{alg:effective_hamiltonians} evolves under $\hat H_j$ for each revaluation period $j \in \{1,\ldots,N\}$.

\begin{algorithm}
    \caption{Imaginary Time Evolution with Continuous Evolutions}
\label{alg:effective_hamiltonians}
    \begin{algorithmic}[h]
        \REQUIRE Rate $r$, Volatility $\sigma$, Maturity $T$, Discretization $\vec{z}$, Revaluation epochs $\ttau_j$, with $N$ number of epochs 
        \ENSURE Option price $u(0, \vec{z})$
        \STATE Prepare initial state $|\psi(\tau)\rangle \leftarrow u(0, \vec{z})$
        \STATE Set current time $\tau \leftarrow 0$ (Imaginary time)
        \FOR{$j = 1,2,\ldots,N$}
            \STATE Construct Evolution Hamiltonian $\hat{H}_{j}$ from \eqref{eq:matrix_HM2}
            \STATE Evolve $|\psi(\tau)\rangle$ to $|\psi(\ttau_j)\rangle$ using $\hat{H}_{j}$ via chosen method (here: \texttt{VarQITE})
            \STATE $|\psi(\tau)\rangle \gets |\psi(\ttau_j)\rangle$
            \STATE $\tau \gets \ttau_j$
        \ENDFOR
        \STATE Extract and adjust price from final state: 
        $u(\tau_N, \vec{z}) \gets \frac{e^{-rT}}{\langle 0| \psi(\tau_N) \rangle} \cdot |\psi(\tau_{N})\rangle$
        \RETURN $u(\tau_N, \vec{z})$
    \end{algorithmic}
\end{algorithm}

\subsection{Variational Approach for Imaginary Time Evolution}

We previously mentioned that the direct implementation of the evolution under $\hat{H}(\tau)
$ is generally not feasible due to the non-unitary nature of the exponential operator. To address this, we employ a variational approach, specifically the Variational Quantum imaginary time evolution (\texttt{VarQITE}) algorithm \cite{McArdle2019}.

The central idea of \texttt{VarQITE} is to replace the exact, non-unitary evolution with a purely unitary, hardware-compatible parameterized circuit, known as a Variational Quantum Circuit (VQC). The target state $\ket{\psi(\tau)}$ is approximated by the parameterized state $\ket{\phi(\vec{\theta}(\tau))}$, where $\vec{\theta}(\tau)$ is a vector of trainable
 circuit parameters. Consequently, the problem of quantum state evolution is transformed into a classical optimization task: determining the optimal evolution of the parameters $\vec{\theta}(\tau)$. This optimization is achieved by minimizing the distance between the exact time derivative of the quantum state and the time derivative generated by the ansatz. McLachlan's Variational Principle (MVP) is typically employed for this minimization, ensuring that the approximate state $\ket{\phi(\vec{\theta}(\tau))}$ closely tracks the true evolutionary path. Application of the MVP leads to a system of linear equations,
$\sum_j A_{i,j}\dot{\theta}_j = -C_i$, where $A_{i,j} = \Re \left( \frac{\partial \bra{\phi(\vec{\theta}(\tau ))}}{\partial \theta_i } \frac{\partial \ket{\phi(\vec{\theta}(\tau ))}}{\partial \theta_j } \right)$, and $C_{i} = \Re \left( \frac{\partial \bra{\phi(\vec{\theta}(\tau ))}}{\partial \theta_i } \hat{H} \ket{\phi(\vec{\theta}(\tau ))} \right)$.
\newline
More details on the \texttt{VarQITE} method can be found in McArdle et al.~\cite{McArdle2019}.

The Variational Quantum imaginary time evolution method, being inherently variational, requires a critical choice of ansatz, namely the trainable parameterized quantum circuit used for the evolution, for its success and efficiency. We rigorously tested multiple ansatz structures to optimize performance for our problem. We decided to exclusively employ the Ry and CRy gates in our final ansatz designs. The primary motivation for this selection is the requirement that the computed quantum state's amplitudes remain real for the extraction of the normalized evolution vector. Using only these gates (and no gates that can introduce complex phases) helps ensure this condition is met. We explored several ansatz constructions during our optimization phase. The construction of the ansatz model used in this work is found in Algorithm~\ref{alg:ansatz}. 

\begin{algorithm}[h]
    \caption{Construction of a layered ansatz}
    \label{alg:ansatz}
    \begin{algorithmic}[1]
        \REQUIRE $n$ qubits, $p$ parameters $\vec{\theta}=(\theta_1,\ldots,\theta_p)$, $k=0$, control qubit $q_c = 0$
        \STATE Prepare $U(\vec{\theta}) = |0\rangle^{\otimes n}$ and apply $H$ on all qubits
        \WHILE{$k < p$}
            \STATE Add $\mathrm{CR}_y$ gate to $U(\vec{\theta})$ parameterized by $\theta_k$, with control $q_c$ and target \\ $q_t =  (q_c + 1) \mod n$; update $k \leftarrow k + 1$, $q_c \leftarrow q_t$
            \STATE \textbf{IF} $q_t = 0$, apply a layer of $R_y$ gates to all $n$ qubits, parameterized by $(\theta_k, \ldots, \theta_{\max (k+n-1, p-1)})$ ; update $k \leftarrow k + n$
        \ENDWHILE
        \STATE \textbf{return} $U(\vec{\theta})$
    \end{algorithmic}
\end{algorithm}

\subsubsection*{Initial State Preparation and Parameter Initialization}
To effectively train the variational circuit used in the imaginary time evolution, we first require a high-quality reference state, $u_{\text{initial}} \equiv |\psi_{\text{initial}}\rangle.$
This state is prepared using the generic quantum state preparation algorithm of Iten \textit{et al.}~\cite{Iten2016}, which is implemented in the \texttt{Qiskit} quantum computing framework. The routine produces a circuit that prepares $|\psi_{\text{initial}}\rangle$ from the computational basis state $|0\rangle^{\otimes n}$.

The prepared state serves as the training target for the chosen $R_y$/CR$_y$ variational ansatz. The objective of this pre–training stage is to determine a set of parameters $\theta_{\text{initial}}$ such that the ansatz reproduces the target state,
$
U(\theta_{\text{initial}})|0\rangle \approx |\psi_{\text{initial}}\rangle.
$
We quantify the quality of the approximation through the quantum state fidelity $F(\theta)
|\langle 0|U^\dagger(\theta)|\psi_{\text{initial}}\rangle|^2$. The training uses measurement-based overlap estimation instead of directly comparing state vectors. We minimize $C(\theta) = 1 - P_{0^n}(\theta)$ where $P_{0^n}(\theta) = |\langle 0|U^\dagger(\theta)|\psi_{\text{initial}}\rangle|^2$. $P_{0^n}(\theta)$ is estimated by repeated quantum measurements, and a classical optimizer (here, \texttt{Scipy's} BFGS) finds $\theta_{\text{initial}} = \arg\min_\theta C(\theta)$ (Algorithm~\ref{alg:init_params}). The optimized parameters $\theta_{\text{initial}}$ are used as the starting point for the imaginary time evolution optimization loop (VarQITE). 

\begin{algorithm}[h]
    \caption{Optimization of Initial Ansatz Parameters}
    \label{alg:init_params}
    \begin{algorithmic}[1]
        \REQUIRE Variational ansatz $U(\theta)$, prepared state $|\psi_{\text{initial}}\rangle$, optional initial parameters $\theta_0$
        \ENSURE Optimized parameters $\theta_{\text{initial}}$ and final cost $C(\theta_{\text{initial}})$
        
        \STATE Construct an $n$-qubit quantum circuit
        \STATE Prepare the state $|\psi_{\text{initial}}\rangle$ \cite{Iten2016}
        \STATE Apply the inverse ansatz $U^\dagger(\theta)$
        \STATE Measure all qubits in the computational basis
        
        \IF{$\theta_0$ not provided}
            \STATE Initialize $\theta_0$ randomly
        \ENDIF
        
        \STATE Define the cost function
        $C(\theta) = 1 - P_{0^n}(\theta)$
        
        \STATE Estimate $P_{0^n}(\theta)$ using repeated circuit sampling
        
        \STATE Use a classical minimization routine to compute
        $
        \theta_{\text{initial}}
        =
        \arg\min_{\theta} C(\theta)
        $
        
        \RETURN $\theta_{\text{initial}},\; C(\theta_{\text{initial}})$
    \end{algorithmic}
\end{algorithm}

\section{Results}\label{seq:res}
This section presents a numerical comparison of Algorithm \ref{alg:sequential_evolution}, including an exact simulation of the time evolution, and Algorithm \ref{alg:effective_hamiltonians} against a highly resolved Monte Carlo (MC) simulation benchmark. The analysis focuses on the fidelity of the calculated option price $u(0, z)$ as a function of maturity time ($T$) and the level of domain discretization (qubit count). Specifically, the variable $z$ ranges from $z_{\min} = 0$ to $z_{\max} = 2.5$, defining a spatial domain discretized into $2^N$ elements, where $N$ is the number of qubits used. Although multiple ansätze have been tested, the results showcased here are using the configuration as described in Algorithm~\ref{alg:ansatz}, with 100 parameters for both 4 and 8 qubits configuration. All ansätze's initial parameters have been trained using the routine described in Algorithm~\ref{alg:init_params}, with a distance to the true initial state of order less than $10^{-6}$. For tests on quantum simulators, we have used 10000 shots per execution. 

\subsection{Algorithm \ref{alg:sequential_evolution} Performance Analysis}

\begin{figure}
     \centering
     \begin{subfigure}[h]{0.48\textwidth}
         \centering
         \includegraphics[width=\textwidth]{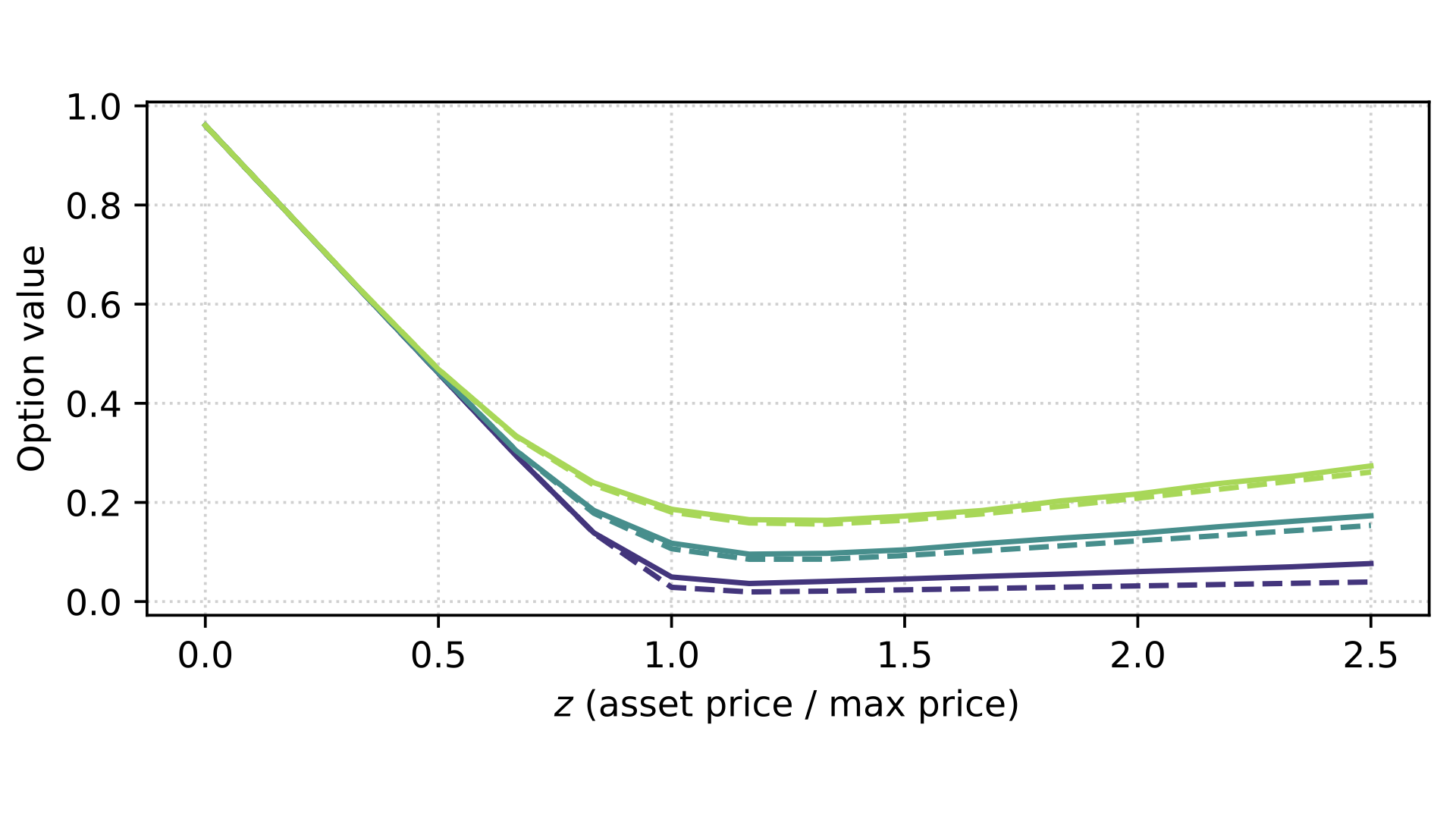}
       \phantomcaption
       \label{fig:method1_t2_q4}
     \end{subfigure}
     \hfill
     \begin{subfigure}[h]{0.48\textwidth}
         \centering
         \includegraphics[width=\textwidth]{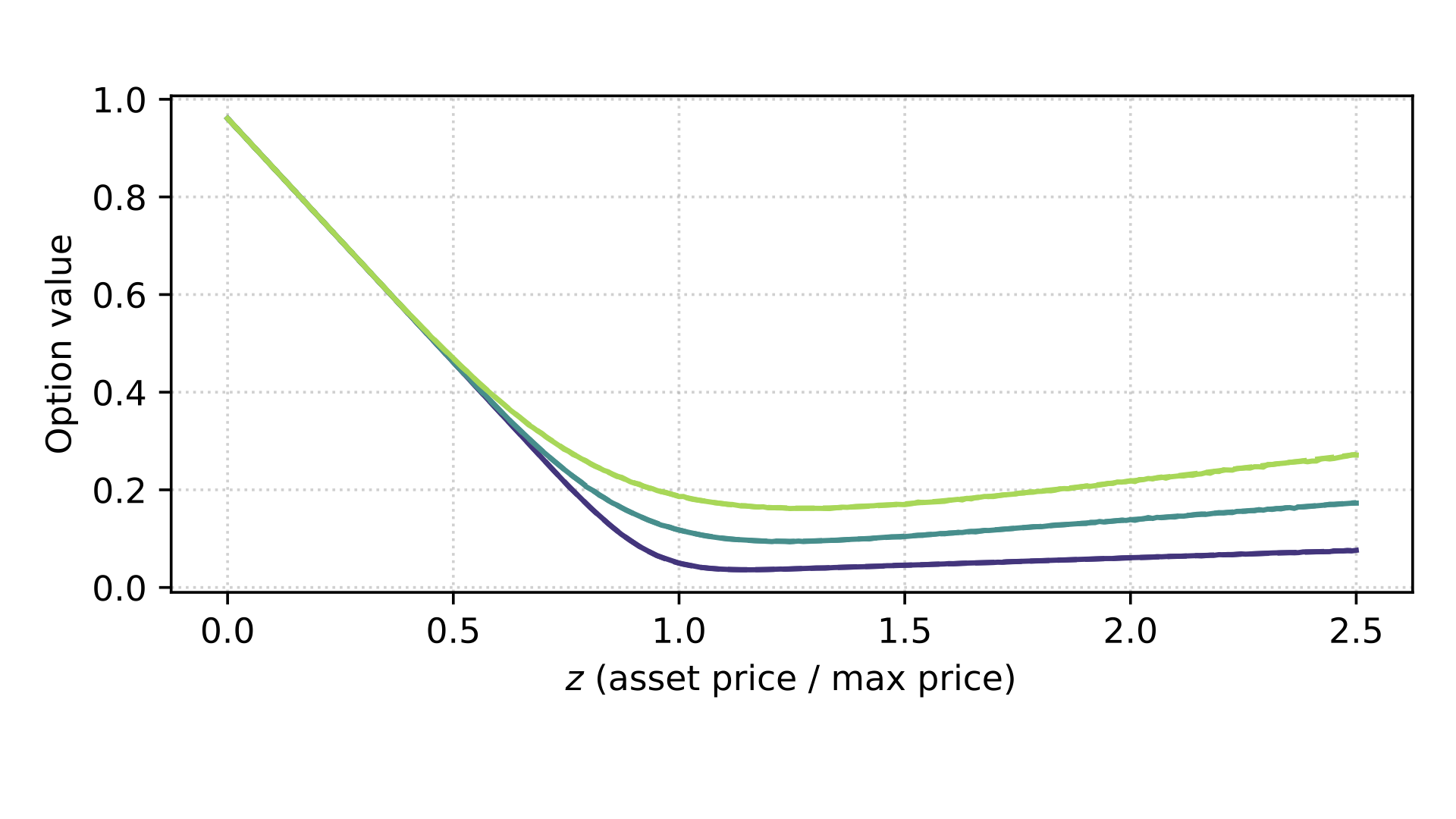}
         \phantomcaption
         \label{fig:method1_t2_q8}
     \end{subfigure}
    \begin{subfigure}[h]{0.7\textwidth}
        \vspace{-25pt}
         \centering
         \includegraphics[width=\textwidth]{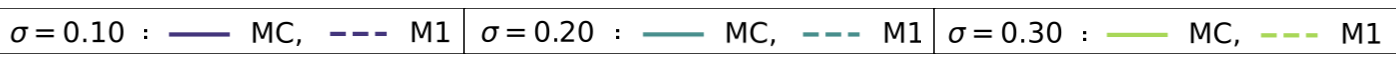}
     \end{subfigure}
     \vspace{-10pt}
        \caption{Algorithm \ref{alg:sequential_evolution} compared to MC for $T=2$ years using 4 qubits (left) and 8 qubits (right). A clear difference in the first plot indicates discretization error; finer discretization improves agreement with the benchmark.}
\end{figure}

The initial application of Algorithm \ref{alg:sequential_evolution} at a short-term maturity of $T=2$ years reveals a strong dependence of solution fidelity on the level of computational domain discretization. Using a minimal 4-qubit configuration, corresponding to $2^4=16$ basis states, the resulting option price exhibits a noticeable divergence from the reference MC benchmark, as depicted in Figure \ref{fig:method1_t2_q4}. This discrepancy is indicative of a substantial discretization error. Upon increasing the system size to 8 qubits, the spatial resolution is refined exponentially. This enhancement leads to a significant improvement in the method's fidelity, achieving closer agreement with the MC reference across the entire asset price domain, as confirmed by the data presented in Figure \ref{fig:method1_t2_q8}.

\begin{figure}
     \centering
     \begin{subfigure}[h]{0.48\textwidth}
         \centering
         \includegraphics[width=\textwidth]{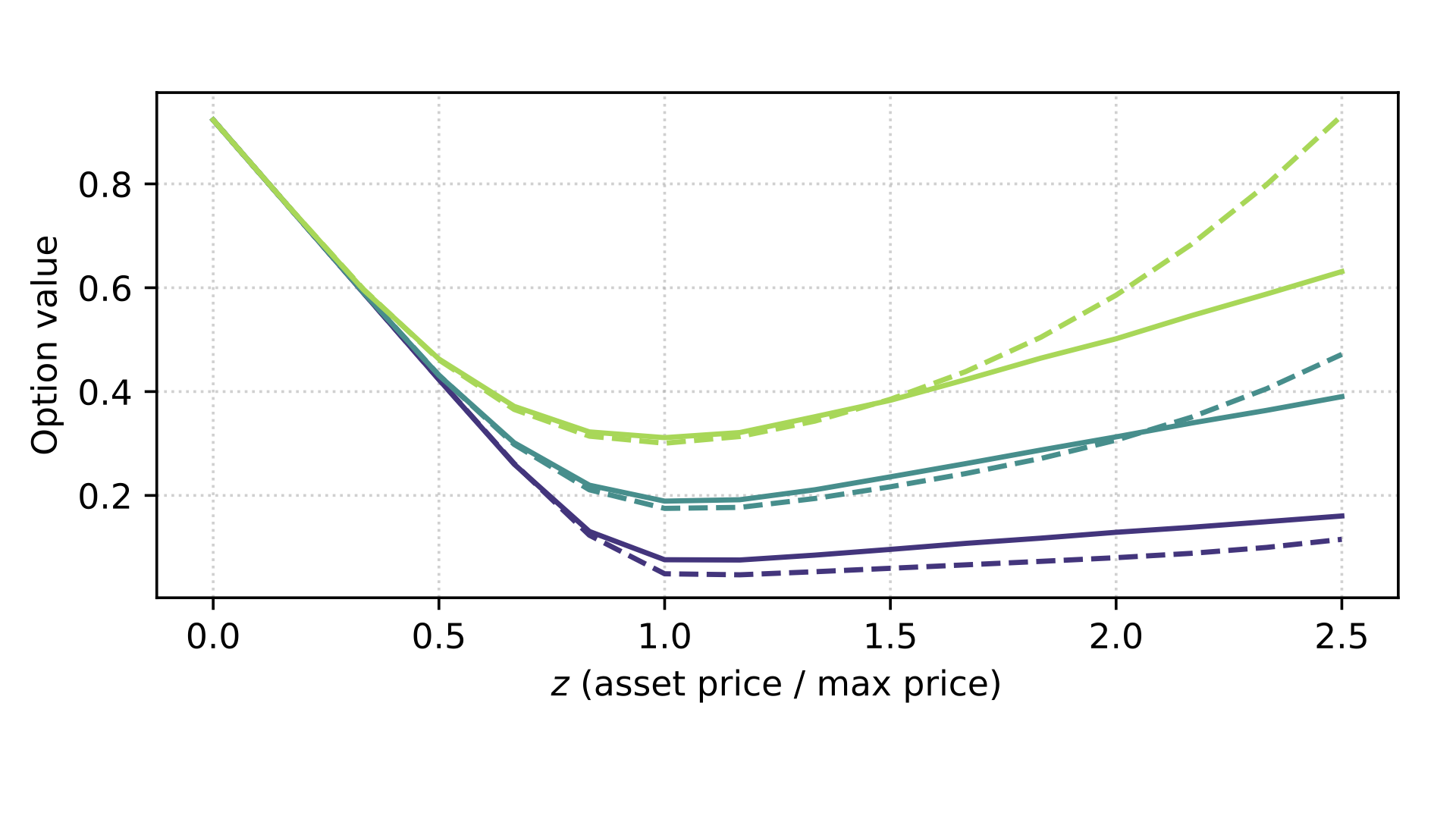}
         \phantomcaption
         \label{fig:method1_t4_q4}
     \end{subfigure}
     \hfill
     \begin{subfigure}[h]{0.48\textwidth}
         \centering
         \includegraphics[width=\textwidth]{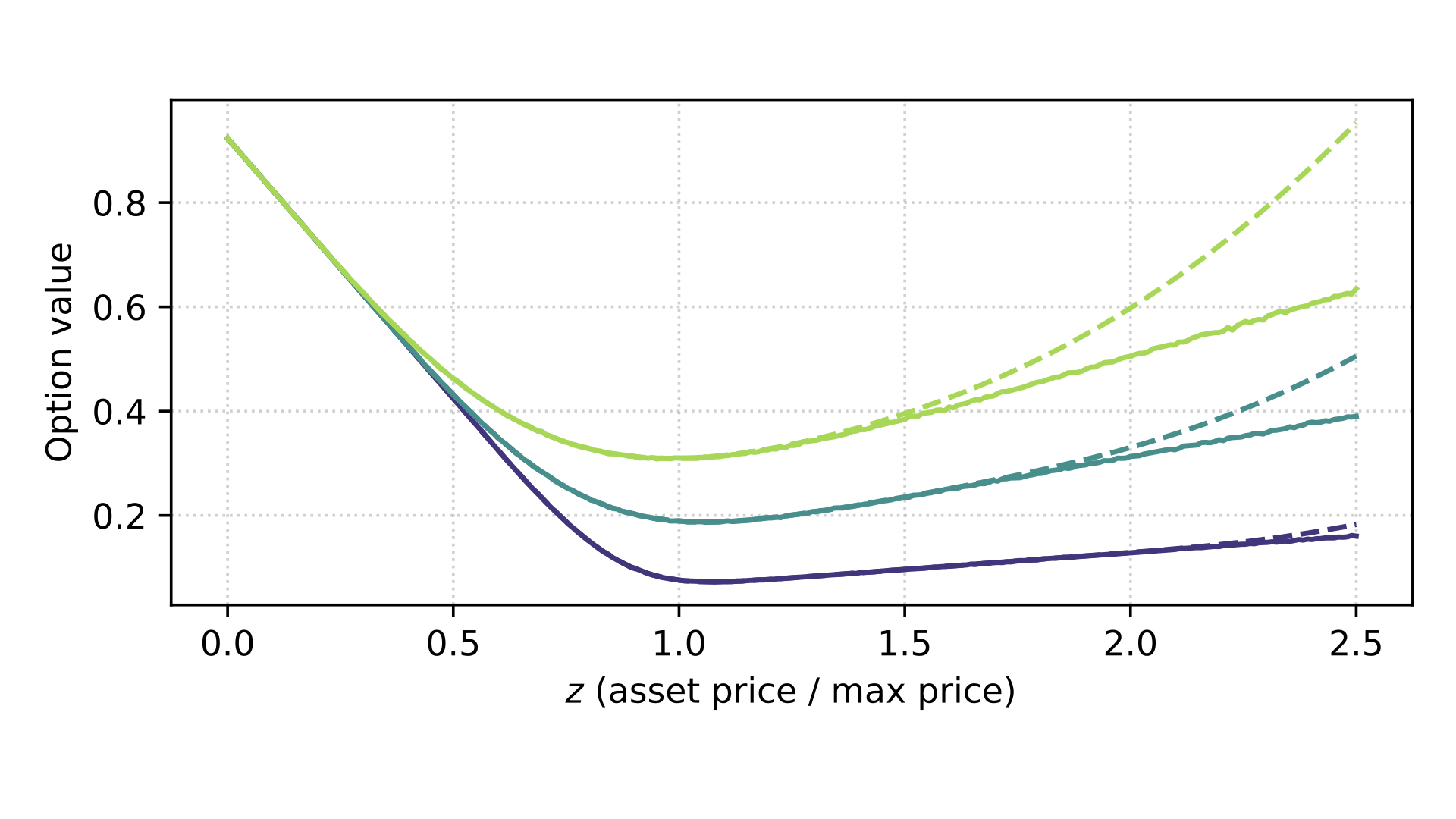}
         \phantomcaption
         \label{fig:method1_t4_q8}
     \end{subfigure}
    \begin{subfigure}[h]{0.7\textwidth}
        \vspace{-25pt}
         \centering
         \includegraphics[width=\textwidth]{plots/M1.png}
     \end{subfigure}
     \vspace{-10pt}
        \caption{Algorithm \ref{alg:sequential_evolution} compared to MC for $T=4$ years using 4 qubits (left) and 8 qubits (right). Error seems to accumulate through the years.}
\end{figure}

Extending the time horizon to a maturity of $T=4$ years introduces increased computational challenges, primarily due to the accentuation of accumulated errors stemming from the sequential, iterative application of the continuous Evolution Hamiltonian ($\hat{H}_C$) and the Jump Hamiltonian ($\hat{H}_J$) in Algorithm \ref{alg:sequential_evolution}. The 4-qubit simulation manifests a significant discrepancy relative to the MC benchmark (Figure \ref{fig:method1_t4_q4}), highlighting the instability of the sequential evolution over prolonged evolution periods. While scaling to 8 qubits mitigates the error, yielding a result demonstrably superior to the 4-qubit configuration, a non-negligible residual divergence from the MC price is still observed (Figure \ref{fig:method1_t4_q8}). A key contributing factor identified is the volatility ($\sigma$). Elevated volatility increases the spectral spread of the Hamiltonian, consequently complicating the landscape of the evolved quantum state. This increased complexity makes convergence more difficult for the fixed-depth ansatz, particularly when integrated with the noise induced by the repeated jump operations.

\subsection{Algorithm \ref{alg:effective_hamiltonians} Performance Analysis}

Algorithm \ref{alg:effective_hamiltonians}, which fundamentally restructures the problem by leveraging pre-calculated effective Hamiltonians $\hat{H}_j$, $j \in \{1,\ldots,N-1\}$ for each discrete period, demonstrates markedly superior computational stability and accuracy than Algorithm \ref{alg:sequential_evolution} when benchmarked against MC simulation. Here again, we look at an exact simulation of the time evolution. This enhancement in the results is achieved by consolidating the complex dynamics, including the jump effects, into a single, period-specific operator, thereby minimizing the cumulative errors associated with repeated sequential operations. It is important to note, however, that the implementation of the complex, effective operators within $\hat{H}_i$ necessarily mandated the utilization of auxiliary qubits, which impacts the total quantum resource overhead.

\subsubsection{Comparison to Monte Carlo for 2 Years Maturity}
\begin{figure}
     \centering
     \begin{subfigure}[h]{0.48\textwidth}
         \centering
         \includegraphics[width=\textwidth]{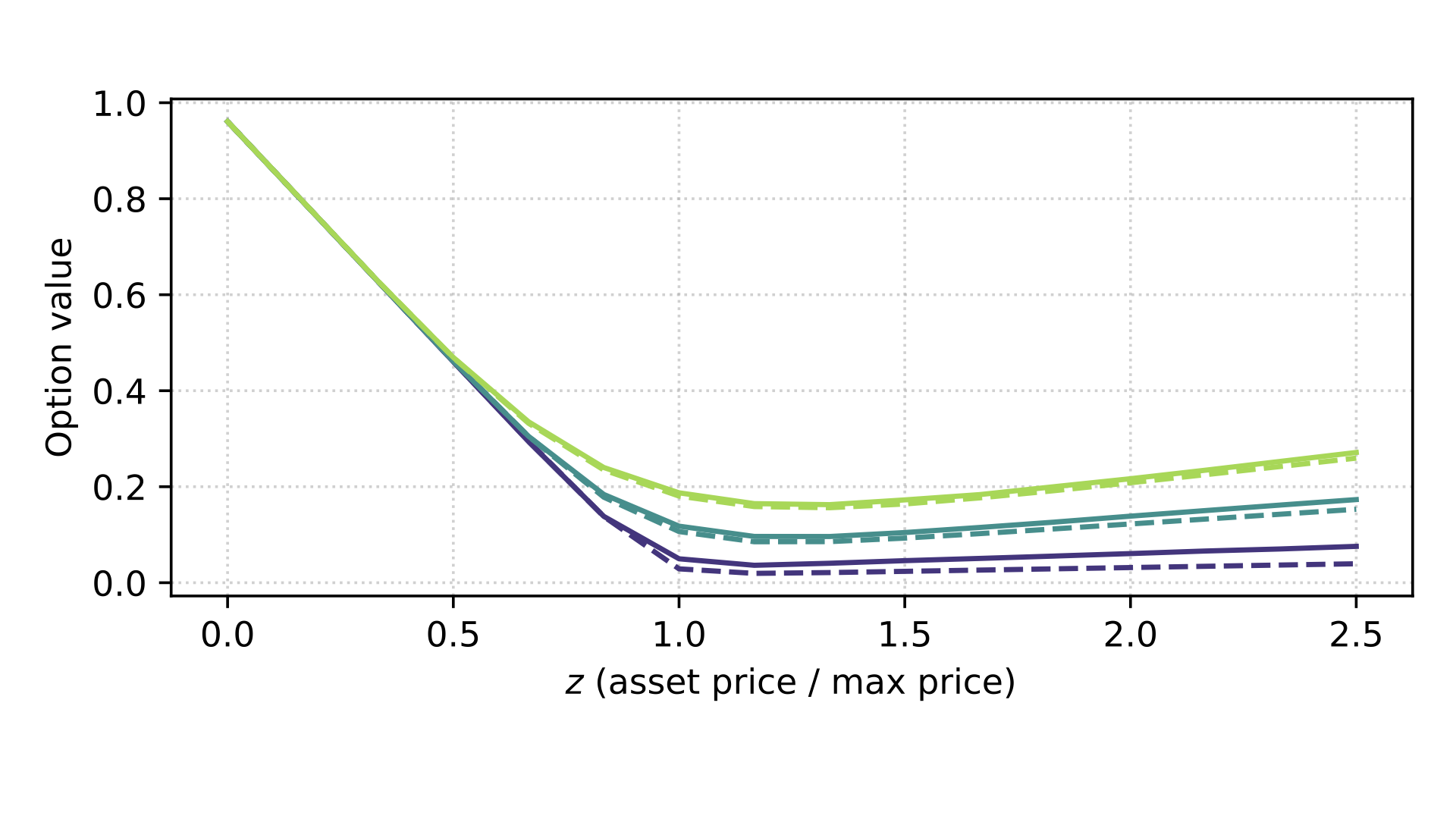}
         \phantomcaption
         \label{fig:method2_t2_q4}
     \end{subfigure}
     \hfill
     \begin{subfigure}[h]{0.48\textwidth}
         \centering
         \includegraphics[width=\textwidth]{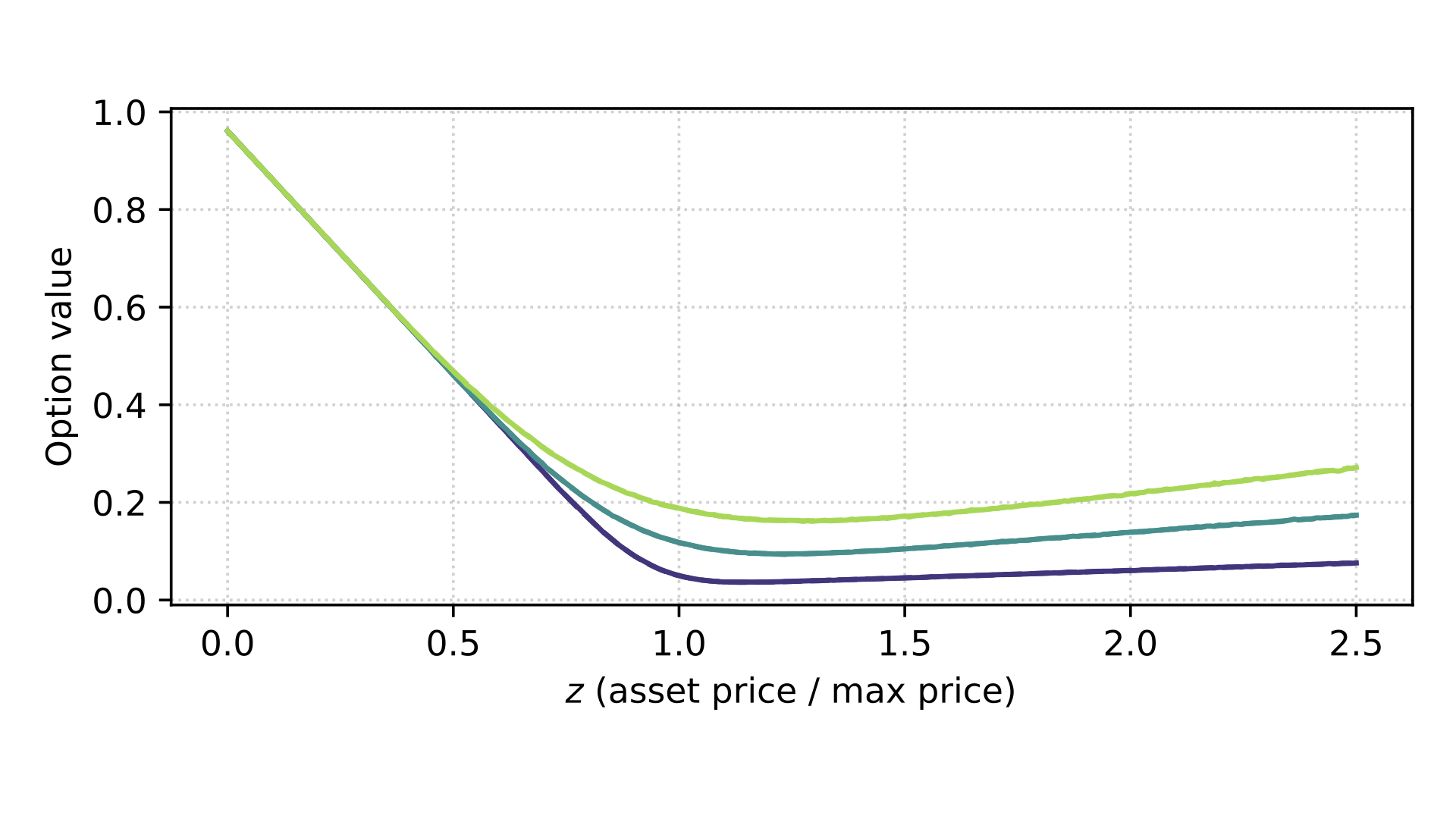}
         \phantomcaption
         \label{fig:method2_t2_q8}
     \end{subfigure}
    \begin{subfigure}[h]{0.7\textwidth}
        \vspace{-25pt}
         \centering
         \includegraphics[width=\textwidth]{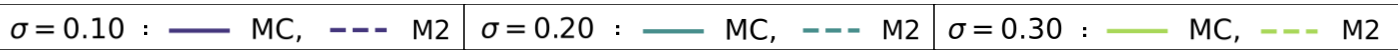}
     \end{subfigure}
     \vspace{-10pt}
        \caption{Algorithm \ref{alg:effective_hamiltonians} compared to MC for $T=2$ years using 4 qubits (left) and 8 qubits (right). A clear difference in the first plot indicates discretization error; finer discretization improves agreement with the benchmark.}
\end{figure}

For the $T=2$ year maturity, Algorithm \ref{alg:effective_hamiltonians} exhibits high fidelity across both discretization levels. The 4-qubit configuration, augmented by the required auxiliary qubits, yields results in strong quantitative agreement with the MC benchmark (Figure \ref{fig:method2_t2_q4}). Using the 8-qubit configuration (Figure \ref{fig:method2_t2_q8}), the method performs exceptionally well, with the calculated option price maintaining high fidelity to the MC reference across the entirety of the underlying asset price domain.

\subsubsection{Comparison to Monte Carlo for 4 Years Maturity}
\begin{figure}
     \centering
     \begin{subfigure}[h]{0.48\textwidth}
         \centering
         \includegraphics[width=\textwidth]{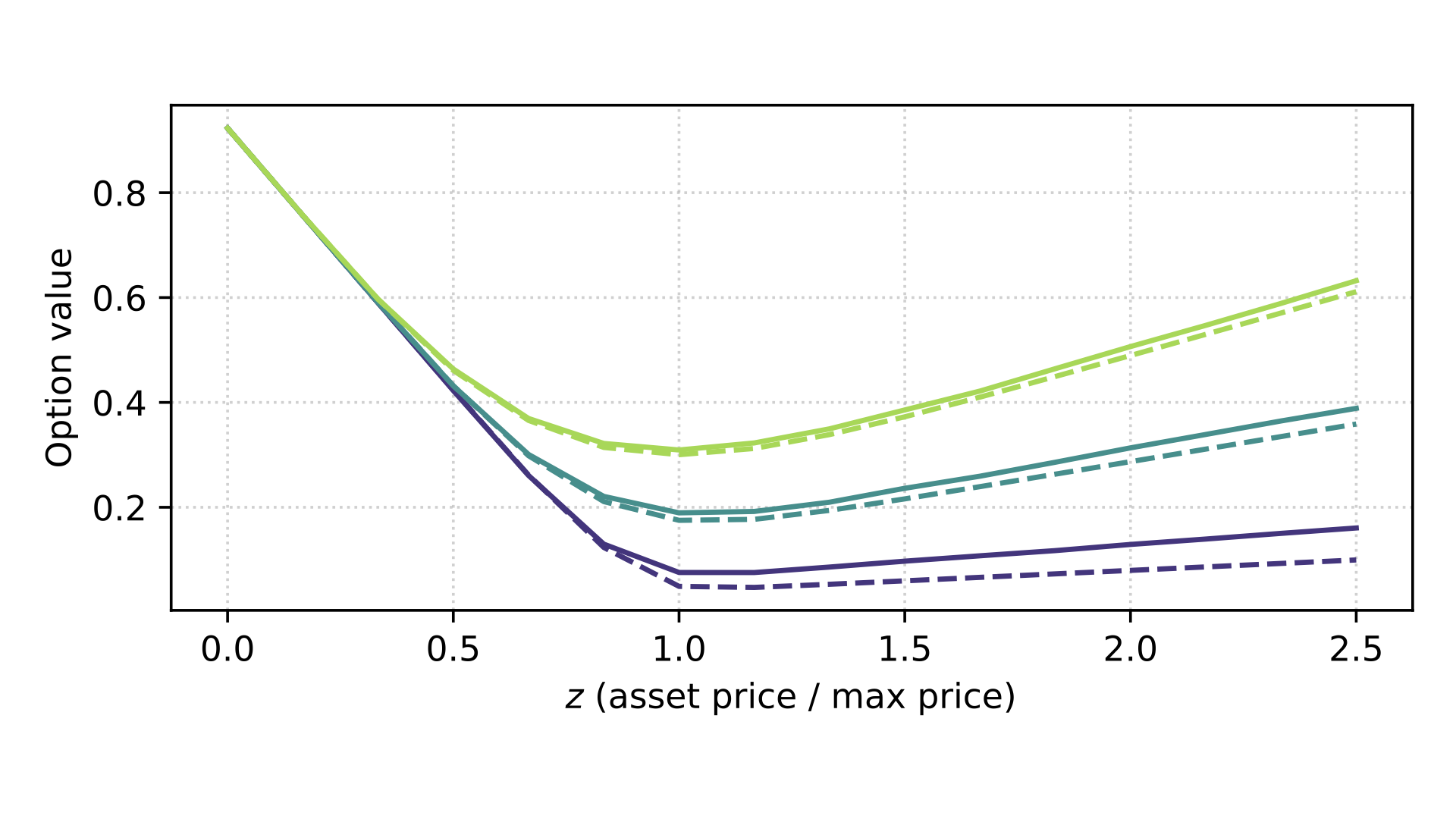}
        \phantomcaption
        \label{fig:method2_t4_q4}
     \end{subfigure}
     \hfill
     \begin{subfigure}[h]{0.48\textwidth}
         \centering
         \includegraphics[width=\textwidth]{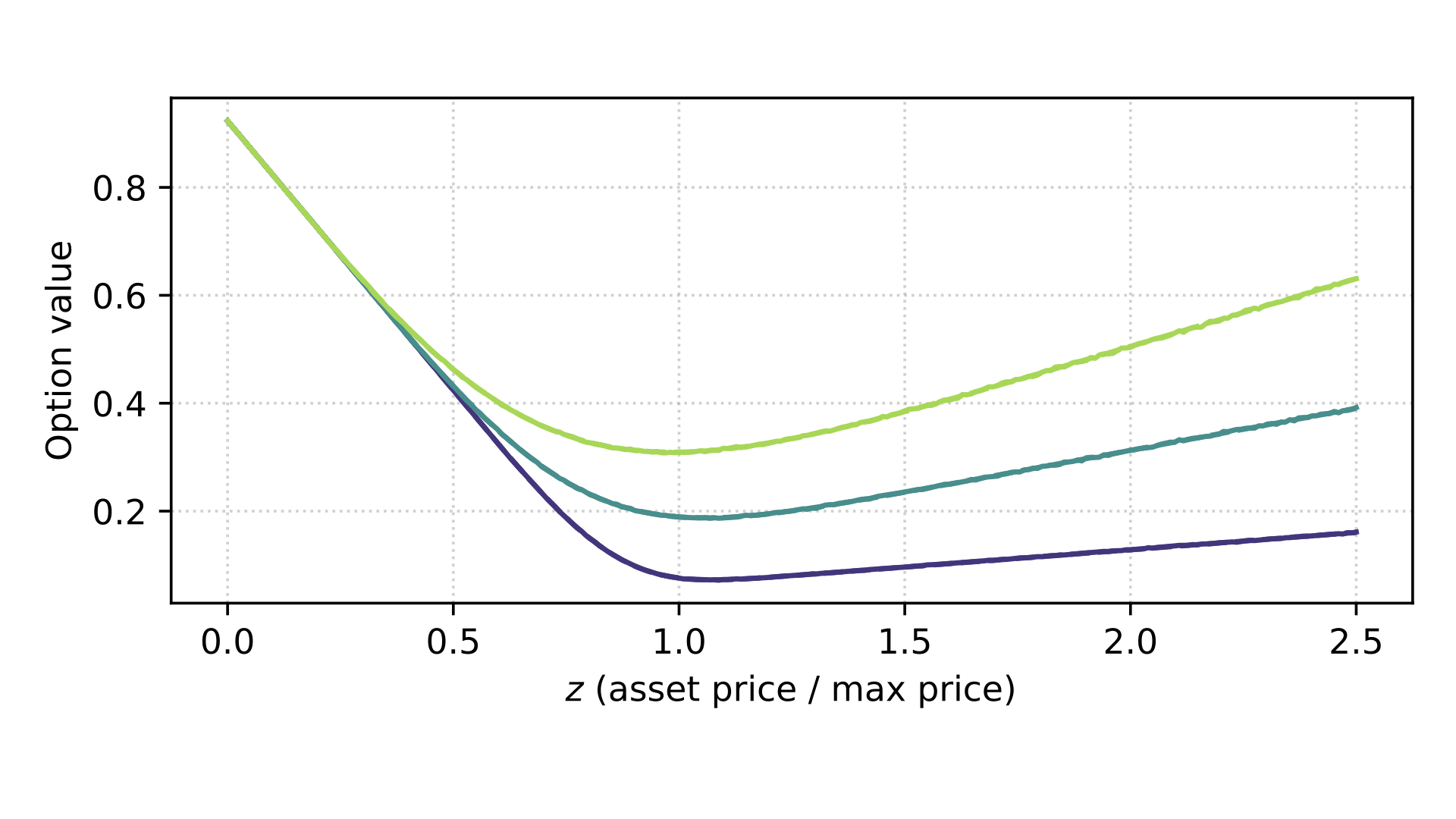}
         \phantomcaption
         \label{fig:method2_t4_q8}
     \end{subfigure}
    \begin{subfigure}[h]{0.7\textwidth}
        \vspace{-25pt}
         \centering
         \includegraphics[width=\textwidth]{plots/M1.png}
     \end{subfigure}
     \vspace{-10pt}
        \caption{Algorithm \ref{alg:effective_hamiltonians} compared to MC for $T=4$ years using 4 qubits (left) and 8 qubits (right).}
\end{figure}

The robustness of Algorithm \ref{alg:effective_hamiltonians} is further validated when extending the maturity to $T=4$ years. The 4-qubit result, while highly accurate, begins to manifest a small deviation from the MC reference (Figure \ref{fig:method2_t4_q4}). This minor divergence is hypothesized to be a consequence of the variational ansatz's limited expressivity over the extended imaginary time path, necessitating an even deeper circuit to fully capture the complexity introduced by the four time periods. Crucially, the 8-qubit simulation maintains its highly accurate and robust performance (Figure \ref{fig:method2_t4_q8}), demonstrating that Algorithm \ref{alg:effective_hamiltonians}'s approach could scale successfully with increased time horizons and problem complexity.

\subsection{Discussion}
The quantitative findings underscore the central role of spatial discretization in the quantum pricing model. The resolution of the underlying asset price domain is determined by the number of qubits ($N$), resulting in an exponential scaling of the grid size ($\Delta x \propto 2^{-N}$). A clear bifurcation in performance is observed: while increasing $N$ from 4 to 8 qubits yielded substantial improvements in accuracy for both algorithms, especially Algorithm \ref{alg:effective_hamiltonians}, the realization of performance gains is not a function of qubit count alone. For highly complex or long-term dynamic regimes, such as the $T=4$ case under elevated volatility, the fidelity is intrinsically linked to the ansatz expressibility. The variational circuit must possess sufficient depth and parameterization to effectively leverage the refined computational grid; otherwise, the high resolution remains computationally underutilized.

\begin{figure}
    \centering
        \includegraphics[width=0.35\textwidth, height=0.3\textwidth]{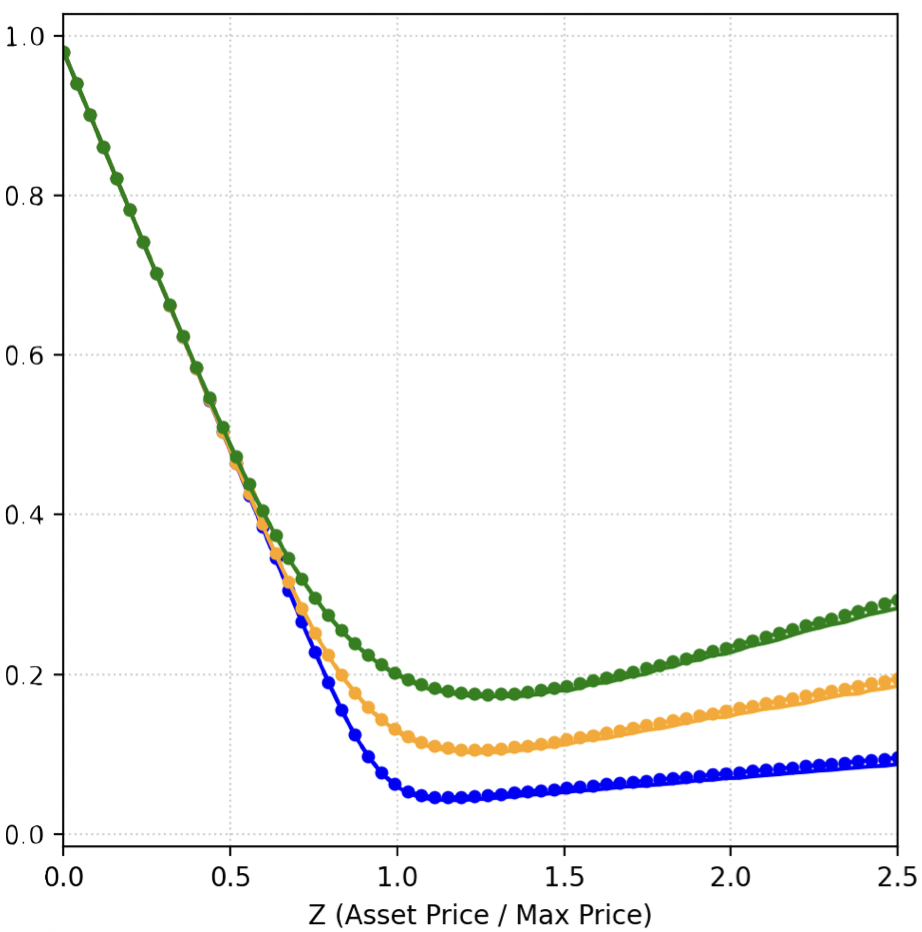}
        \caption{Algorithm \ref{alg:effective_hamiltonians} compared to MC for $T=4$ years using \texttt{VarQITE} to implement Variational imaginary time evolution.}
        \label{fig:varqite_m2}
\end{figure}

When looking at a variational implementation of the imaginary time evolution, we can see in Figure \ref{fig:varqite_m2} that using \texttt{VarQITE} also yields satisfying results. The complexity of the financial product, particularly the revaluation jumps, imposes significant constraints on scaling. Specifically, Algorithm \ref{alg:effective_hamiltonians}'s design, which relies on the Hamiltonians $\hat{H}_i$, necessitates the use of auxiliary quantum registers (auxiliary qubits) to implement the required operators. Consequently, while the $Q$ computational qubits govern the financial resolution, the auxiliary qubits directly increase the total required quantum resource overhead, which is an important consideration for near-term hardware deployment. However, as mentioned earlier, the scaling of the number of qubits follows $\log_2N$, where $N$ is the number of time-periods between revaluations. Furthermore, while the pursuit of higher precision naturally leads to increasing $Q$, future scaling efforts are likely to encounter the intrinsic theoretical limits of variational quantum algorithms, notably the phenomenon of \textit{barren plateaus}, where the optimization gradient vanishes exponentially, thus yielding results of diminishing quality. 

\subsubsection{Hamiltonian Complexity Analysis}

The computational efficiency of both algorithms is predicated on the sparsity and tractability of the Hamiltonian decomposition. As we can see in previous sections, the constructed Hamiltonians are quite sparse. The tensorized Pauli decomposition \cite{Hantzko2024} was utilized to transform the matrix Hamiltonians into a sum of Pauli operators that could be used on quantum hardware. The resulting number of Pauli terms (operators) significantly dictates the necessary depth and overall execution time of the \texttt{VarQITE} circuit. We can see in Table \ref{tab:pauli_terms} that due to the construction of our Hamiltonians in both methods, the number of Pauli terms are greater for method \ref{alg:effective_hamiltonians}. However, since the results of said method are significantly better than those of method \ref{alg:sequential_evolution}, we consider this increase of number of Pauli terms as a necessary trade-off.

\begin{table}[H]
    \centering
    \caption{Number of Pauli Operators in Hamiltonian Decomposition for 4 computational qubits system}
    \label{tab:pauli_terms}
    \begin{tabular}{lccc}
        \toprule
        \textbf{Method} & \textbf{$T=2$} & \textbf{$T=4$} \\
        \midrule
        Algorithm \ref{alg:sequential_evolution} $\hat{H}_{C}$ & 208 & 208 \\
        Algorithm \ref{alg:sequential_evolution} $\hat{H}_{J}$ & 192 & 192 \\
        Algorithm \ref{alg:effective_hamiltonians} $\hat{H}_j$ (mean) & 456 & 2180 \\
        \bottomrule
    \end{tabular}
\end{table}

While \texttt{VarQITE} is a pragmatic solution for NISQ devices, it is subject to the general limitations of variational quantum algorithms. The classical optimization landscape often contains numerous local minima, hindering convergence to the global optimum, particularly with increasing circuit parameter count. Gradient computation can become exponentially vanishing with respect to the number of qubits and circuit depth, severely impeding optimization efficiency. Also, the algorithm remains vulnerable to errors from gate implementation and measurement, necessitating further error mitigation techniques.

\section{Conclusion}
In this work, we developed and analyzed two quantum algorithms to price discretely monitored lookback options within the Black-Scholes framework using imaginary time evolution. Our study demonstrates that the challenges posed by jump conditions in the option pricing PDE, absent in standard European or Asian options, can be effectively addressed within a quantum computational framework.
The sequential evolution approach (Algorithm~\ref{alg:sequential_evolution}) provides a conceptually straightforward method for incorporating jump dynamics, but its performance is sensitive to spatial discretization and suffers from cumulative errors over extended time horizons. In contrast, the effective Hamiltonian approach (Algorithm~\ref{alg:effective_hamiltonians}) consolidates jump effects into period-specific operators, achieving higher fidelity and robustness, albeit at the cost of additional auxiliary qubits and increased Hamiltonian complexity. Numerical simulations show that both methods approach the Monte Carlo benchmarks as the number of qubits and circuit expressibility increase, with Algorithm~\ref{alg:effective_hamiltonians} consistently providing superior accuracy across different maturities and volatility regimes.
Our results highlight key considerations for near-term implementations on NISQ devices, including the trade-off between quantum resource overhead and solution accuracy, as well as limitations inherent to variational algorithms such as barren plateaus and local minima in classical optimization. Looking forward, the transition to fault-tolerant quantum computing could enable non-variational methods that leverage mathematically rigorous error bounds and potentially offer exponential speedups. 
More broadly, this work establishes a concrete framework for the quantum valuation of path-dependent derivatives with discontinuous value functions, hopefully leading toward the pricing of increasingly complex financial products.

\section*{Acknowledgments}
Financial support and in-kind contribution from the Autorité des marchés financiers (AMF Québec) is gratefully acknowledged. 
The authors bear full responsibility for the information contained therein. 
The views and opinions expressed in this paper are those of the authors and do not necessarily reflect the official policy or position of the AMF Québec. 
The support of the Algolab of the Quantum Institute of the University of Sherbrooke is also acknowledged. 
This research was enabled in part by support provided by Calcul Québec (calculquebec.ca) and the Digital Research Alliance of Canada (alliancecan.ca).
The authors acknowledge the use of AI to edit the written text for spelling, grammar, and general style.

\bibliographystyle{siamplain}
\bibliography{references}
\end{document}